\documentclass[a4paper,12pt]{article}

\usepackage{amsmath}%
\usepackage{amsfonts}%
\usepackage{amssymb}%
\usepackage{graphics} 
\usepackage{graphicx} 

\numberwithin{equation}{section}

\newcommand{\norm}[1]{\left\Vert#1\right\Vert}
\newcommand{\abs}[1]{\left\vert#1\right\vert}

\begin{document}
 
\title{On chaotic nature of speech signals}
 
\date {} 
\author{Yu.V. Andreyev$^{*}$, M.V. Koroteev\thanks{Institute of Radioengineering and Electronics of RAS, Moscow, Mokhovaya str. 11/7, e-mail: (Yu.V. Andreyev) - yuwa@cplire.ru, (M.V. Koroteev) - m.koroteev@gmail.com. The article was published in Izv. VUZov. Applied nonlinear dynamics, 2004, v.12, N6, P. 44-59.}}
\maketitle

\begin{abstract}
Various phonemes are considered in terms of nonlinear dynamics. Phase portraits of the signals in the embedded space, correlation dimension estimate and the largest Lyapunov exponent are analyzed. It is shown that the speech signals have comparatively small dimension and the positive largest Lyapunov exponent
\end{abstract}

\section{Introduction} 
\noindent

In the last decades a certain raise of interest to the investigations of speech processes is observed. It is connected both with the practical problems of information compression, stipulated by the increasing role of communications in the society, and with the development of new methods for research of complex systems. 

In the past the basic progress in the investigations of the speech process, especially from the viewpoint of applications, was achieved by means of linear prediction\cite{1,2}. It is based on the fact that the changes of the vocal organs occur at the much longer time scales than the specific period of oscillations for the speech signal. Therefore, at the small time scales the organs of speech may be treated as a stationary system and be replaced by a linear filter, excited by the special signal. Even though this approach is based on the linear prediction, this term must not deceive: it implicitly assumes the nonlinear origin of the speech process. From the one hand, the prediction of the next point of a discrete speech signal is accomplished by means of a linear combination of a number of previous points, but from the other hand, the exciting signal in this model turns out to be a complicated one with the infinite spectrum, i.e., a periodic sequence of pulses for the vocals and a noisy signal for the consonants. Nevertheless, the question of the origin of source generating both the wideband periodic signals and noise-type signals is still slightly discussed.

Apart form the linear prediction, the approach connected to the physical simulations of the speech process was investigated, e.g., for the vocals the nonlinear model\cite{3} was suggested. In this model the vocal cords are represented by virtue of the oscillating masses and the vocal tract, as a double-sided line of transmission. The hardness of connection between the oscillating masses represents the nonlinear element. It is described by piecewise-cubic discontinuous function and the discontinuity corresponds to the glottal stop. When numerically estimating, this function is approximated by a piecewise-linear function. The model is described by the system of differential equations. The solutions of the system for the various sets of parameters turns out to approximate some vocal sounds sufficiently well\cite{3}.

Together with the theory of chaos and of the methods for dynamic systems analysis including the methods for inverse problems of nonlinear dynamics, e.g., reconstruction of attractor from the observations of single variable\cite{4,5,6}, methods of reconstruction of equations for nonlinear systems\cite{7} etc., the works appeared which try to apply these methods for the investigations of speech signals.
Among them it is worth the works\cite{8,9} to be mentioned, in which the transition phenomena in the specific sounds such as monkey and baby cries were investigated. There were discovered in them the double period bifurcations, tori, chaos transition etc.

The observed phenomena point to the certain similarity of the dynamic system of speech producing with the behavior of the artificial chaotic systems. Therefore, it is of some sense to try to reconstruct the unknown, in general, multidimensional dynamic system using the observed speech signal which can be represented in the real situation e.g. by the voltage on the microphone output or to try to estimate its parameters. In \cite{10,11} such  investigations of the speech signal as the reconstruction of phase portrait, Poincare cross-section construction, the estimation of the phase space dimension of the appropriate dynamic system, the largest Lyapunov exponent estimation were done. For the analysis rather short quite stationary fragments of japanese vocal phoneme 'a' were used and the part of the research was accomplished applying the phoneme 'i'.

In our situation the dynamic parameters of the speech signals are investigated with much wider material including some fricative sounds(hissing and sibilant sounds) along with basic vocal sounds. Note that the non-stationarity of the speech processes complicates the application of the methods of nonlinear dynamics. Therefore, the first step consists in the investigation of separate vowels and consonants and then it would be possible to consider the realistic speech signal. Despite the difficulties connected with the non-stationarity it is the non-stationary dynamics that contains the main part of information content of the speech signal. A stationary quasi periodic or noisy signal brings trifling information compared to the nonstationary speech signal. The information content of the speech is stipulated by its non-stationary property being the consequence of continuous changes of the vocal organs while producing the separate phonemes the vocal tract is almost fixed.
  
To confirm the chaotic origin of speech signal the reconstruction of the dynamic system corresponding to the speech signal, 
the estimations of correlation dimension, of the largest Lyapunov exponent have been done. The difficulties appearing while application of methods of nonlinear dynamics to the speech signals and the problem of the obtained results interpretation are discussed.

\section{Reconstruction of the phase space for the speech signals}

When studying time series, the methods such as Lyapunov exponents estimate, correlation dimension estimate, main component analysis are often applied. These methods are also used here but the first problem to be solved prior to application of the methods of nonlinear dynamics is that of the reconstruction of the set, corresponding to the speech signals on the basis of the time series.

The basic material for the analysis is represented by the analog speech signal. In order not to loose the crucial information, the reasonable discretizing frequency value should be specified. One often suppose that the band of the speech signal is sufficiently narrow and for most of applications it can be restricted by the value of the order of 4 kHz, e.g. in the telephone communications the signal is restricted by the band 300-3500 Hz, for the many applications dealing with the speech compression the signal is discretized with the frequency 8 kHz and according to Kotelnikov theorem the signal is restricted in the band 4 kHz. One of the results of the suggested investigation consists in the fact that for some sounds having the noisy part in 3-10 kHz it is not sufficient, though it may be not crucial for the problems of compression and recognition. In our treatment we discretized vowels with the frequency 11,025 kHz and some consonants with 44,1 kHz.

The vowels 'a', 'o', 'i', 'ae', 'u' and the consonants 'sh', 's', 'z' were mostly considered. The sounds were taken both from the real speech and had been recorded separately by males and females. The duration of the recorded sounds was about some seconds but for the analysis the fragments of duration 0.5-1 s. were used. Besides, the stationary parts for each sound were cut from the records. The discretized speech signal (time series) was treated as a signal generated by a nonlinear dynamic system. It is necessary to note that all the investigation methods are well suitable for stationary ergodic systems but it is evident that the vocal tract can not be treated as a stationary system and it is impossible to use the term attractor for this system in the rigorous sense. Therefore, we consider only sufficiently stationary parts of speech signal.  

In addition to the non-stationarity the serious difficulty of application of nonlinear methods to the speech is stipulated by the time multi scaling, i.e. in the signal there exist the oscillations with the different frequency scales. It results that the significant frequencies in the spectrum of the signal seize the range from the hundreds Hz to some kHz differing from the situation with the ideal chaotic systems such as, e.g. Rossler system.
In most cases for the reconstruction of the phase space of a dynamic system the method based on Takens theorem is applied. The essence of the method consists in geometric interpretation of the dynamic system signal in a euclidean space of the prescribed dimension $d$. The time series $\{x_{k}\}$, $k=1,2,\ldots n$ is resampled with the step $\tau$. To obtain the point of the embedding space $R^{d}$, $d$ samples are taken in turn from the resampled time series so that $\vec{x}_{i}=(x_{i},x_{i+\tau},\ldots,x_{i+(\tau-1)d})$\cite{4,6,12,13}. The points of arithmetic space are mapped one-to-one to 
$R^{d}$. The value $\tau$ is usually called lag or time of delay\cite{13,14}. The estimation of the lag is of great interest problem in the literature\cite{13,14} and in two words is formulated as the choosing of 2-4 points on the quasi period of oscillations. However, as it was noted above, in the speech signals there exist the oscillations with the crucially different periods. If the period of the fast oscillations is taken into account, then in the reconstructed $d$-dimensional set the slow dynamics would not be revealed. To allow for this slow motions too large dimension of the space must be chosen. From the other hand, if the slow oscillations are taken into consideration, the fast motions may be lost. Apparently, for the speech signals instead of the uniform resampling (according to Takens) a nonuniform one should be applied when the vectors of $R^{d}$ are chosen to contain both the closest and far elements of the time series involved. Some recommendations concerning these methods can be found in\cite{15}.
 
In our investigations the uniform resamling when embedding the trajectory in $R^{d}$ was used. The application of the very long parts of the trajectories (up to hundreds thousands points) allows to construct 'dense'  sets in $R^{d}$ reducing, in authors` view, unpleasant effects of multi scale dynamics of the signals. The reconstructed points represent the phase trajectory of the dynamic system under consideration and are depicted as the phase portraits of the appropriate speech signals. We also use the sonograms for the analysis of the sounds which are the graphs having the time along $Ox$ and the frequency along $Oy$. In addition, the brightness is proportional to the energy of the frequency in the specific time moment. Thus, the sonogram shows the temporal evolution of the signal. In fig. 1 and 2 the wave forms and the sonograms for the vowel 'a' and consonant 'sh'  are shown.
\begin{figure}
\center{
\includegraphics[width=180pt]{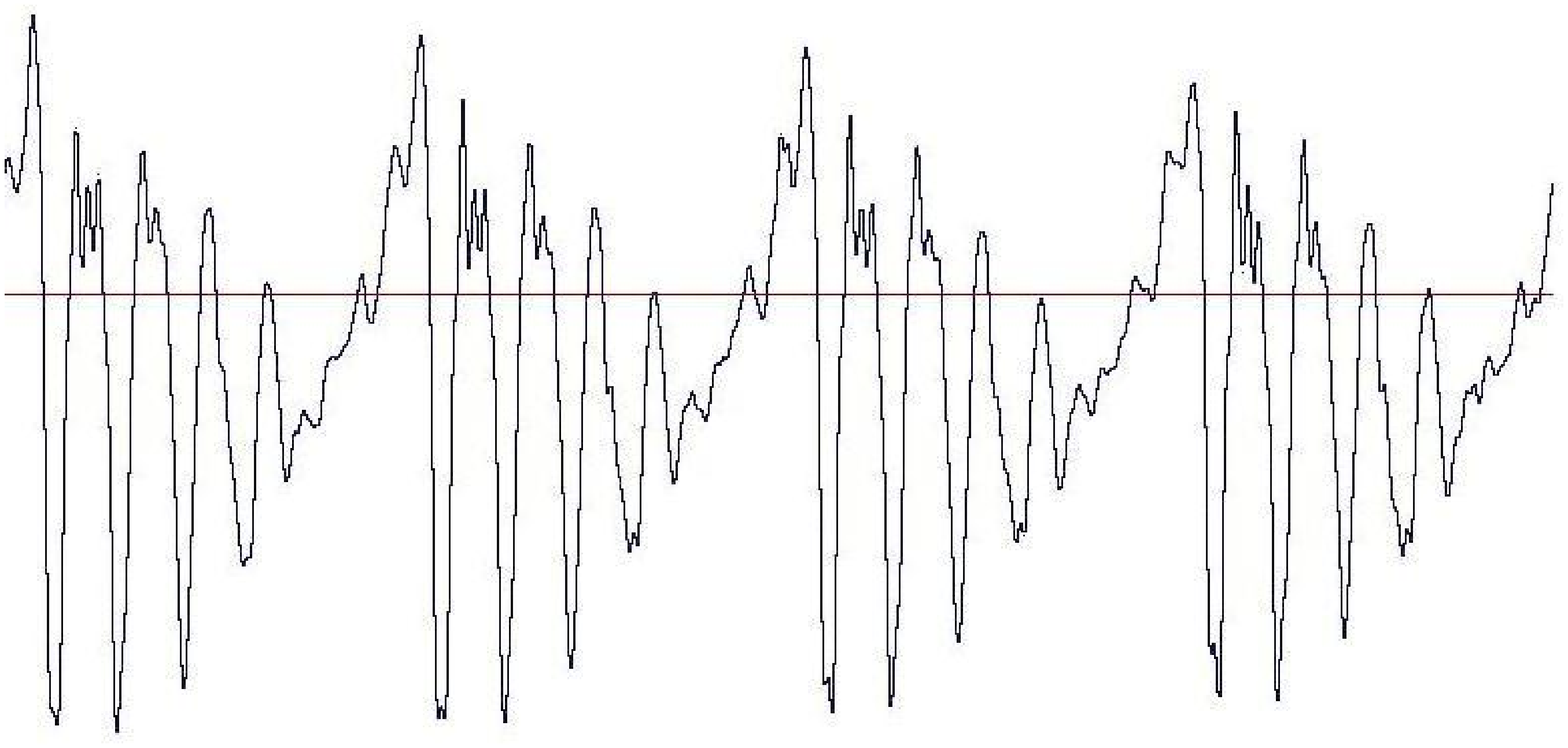}
\includegraphics[width=180pt]{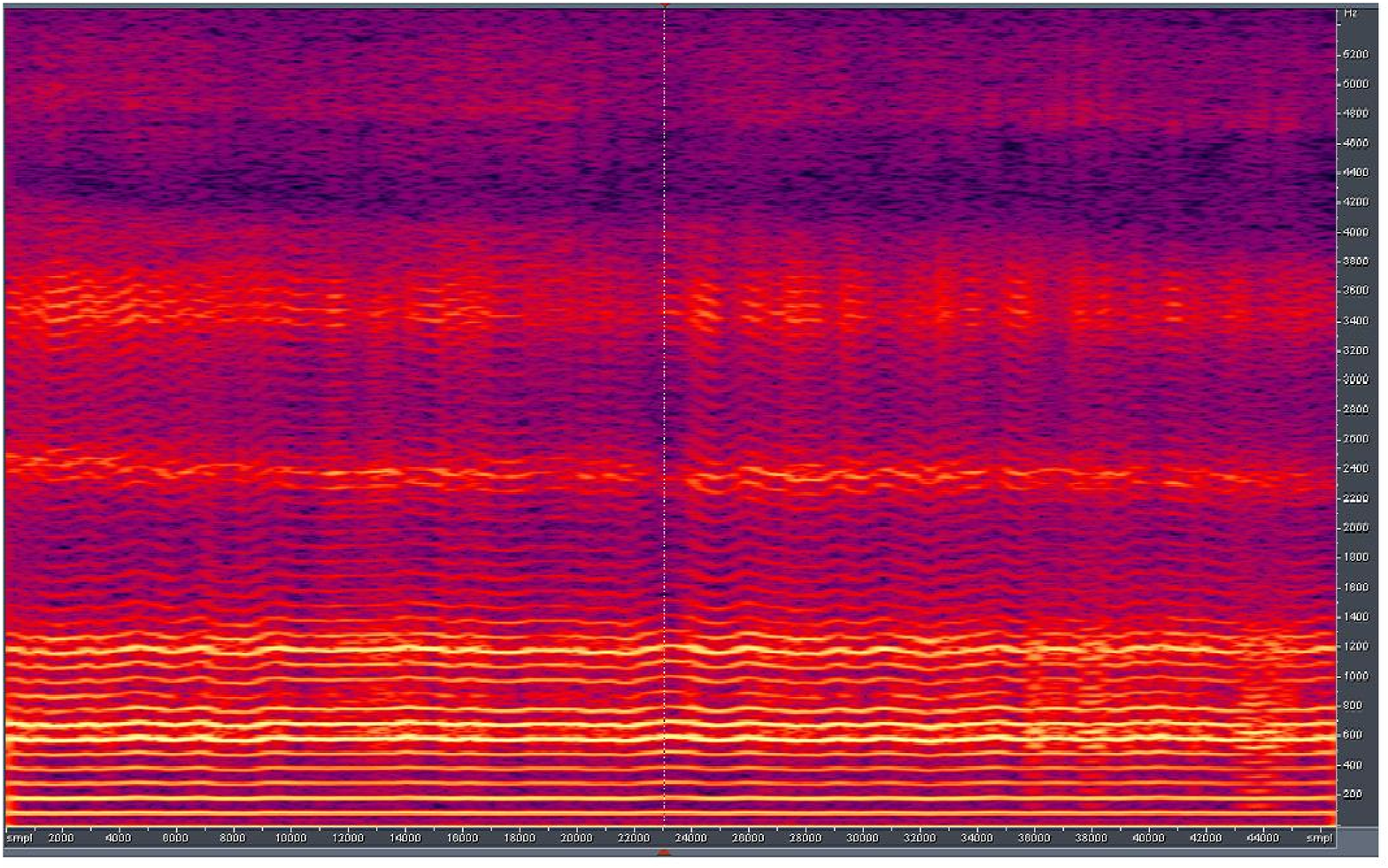}}
\caption{waveform and sonogram for the vowel "a"}
\end{figure}
\begin{figure}
\center{
\includegraphics[width=180pt]{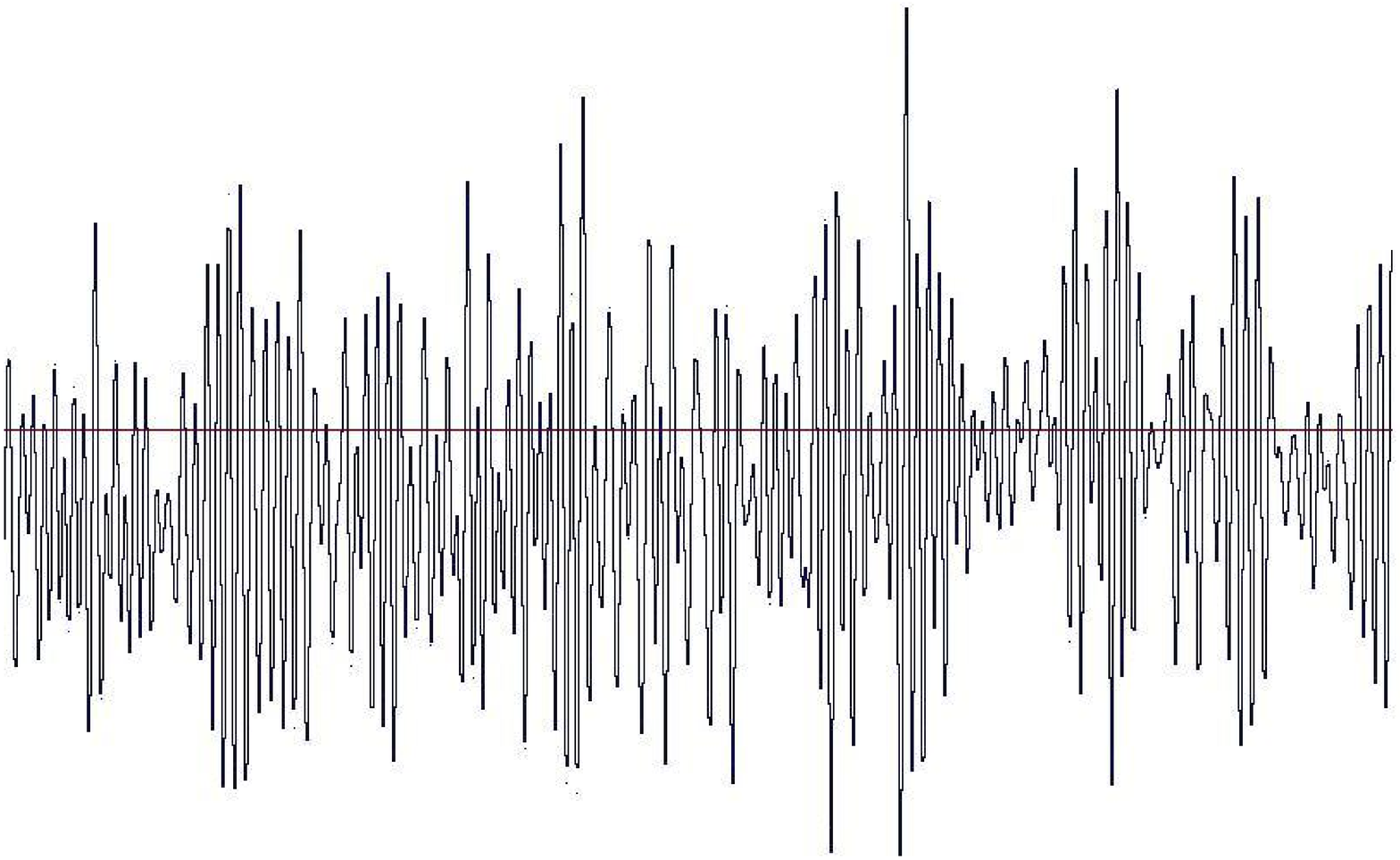}
\includegraphics[width=180pt]{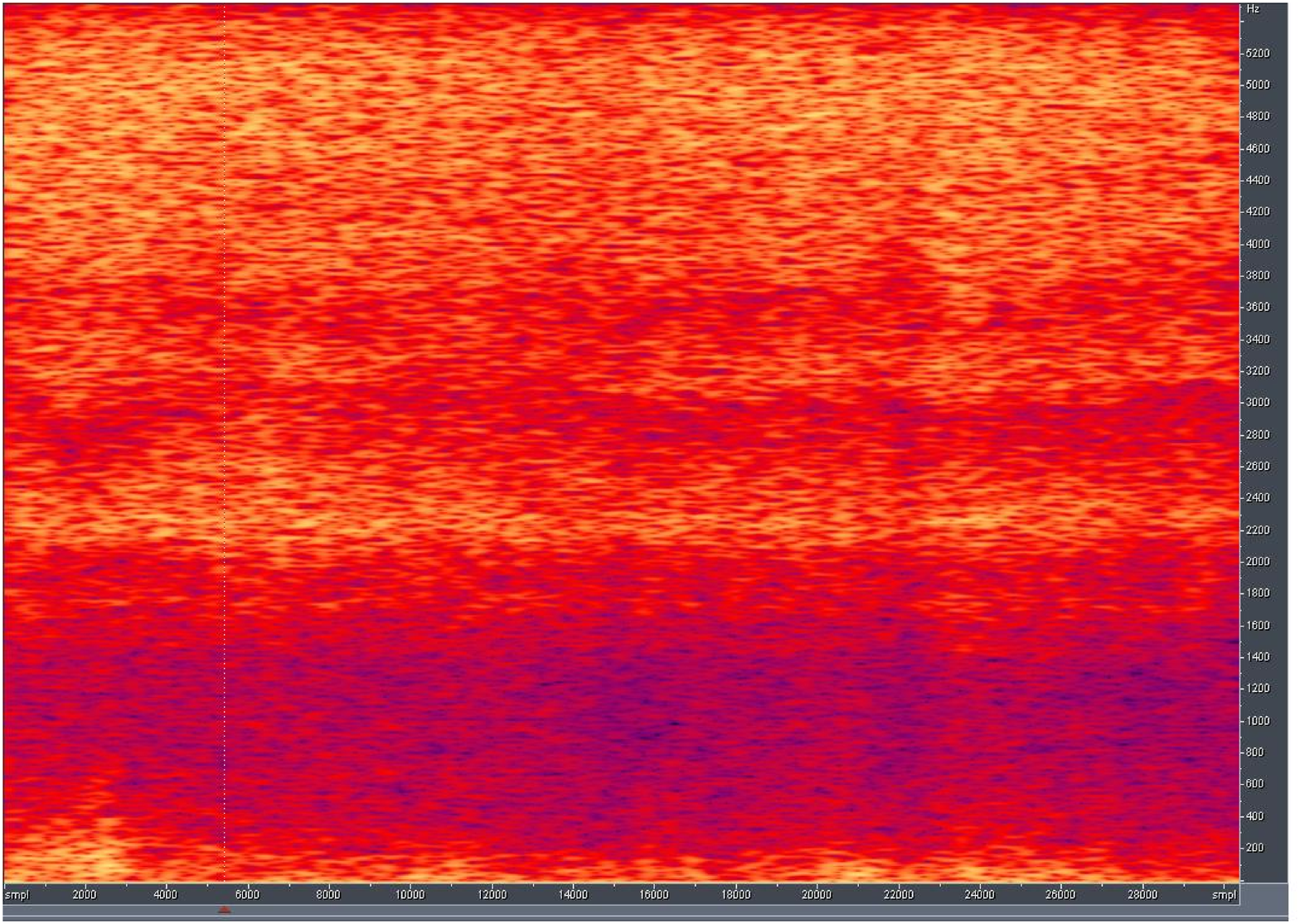}}
\caption{waveform and sonogram for the consonant "sh"}
\end{figure}

It is important to note that the vowels reveal the noticeable quasi-periodic structure while the consonants are of pseodonoise character that can be easily seen both on the wave forms and on the sonograms. According to the example of 'a' sound, the energy of vowels is concentrated in the lower part of the spectrum and the latter has sufficiently noticeable peaks on the fundamental frequency and harmonics that can be seen as the light bands on the sonogram while for 'sh' the energy is approximately uniformly distributed in the spectrum and the very spectrum is similar to that of noise, i.e. it does not contain the visible peaks. The similar properties are also observed for the other consonants under consideration ('s', 'f'). 

When constructing the phase portraits corresponding to the speech sounds, the estimate of the dimension of the space containing the set corresponding to the sound can be done. The Brumhead-King procedure similar to Karhunen-Loeve transform\cite{12} was applied for this estimate. The essence of the method consists in constructing a new orthonormal basis in $R^{d}$ for that set, which is optimal in the sense of mean square error of approximation of this set. The basis is constructed from the eigen vectors of covariance matrix of set vectors and the sum of squared projections of the set on the axes of the new basis, i.e. de facto, the energy corresponding to the specific direction is determined by the eigen values of the covariance matrix. It follows that if the set can be embedded in the subspace of $R^{d}$ then some eigen values will be equal to zero or vanish if the initial data were subjected to the impact of noise. In the numerical experiment the above procedure was accomplished as follows. A matrix $X = (\vec{x_{1}}, \vec{x_{2}},\ldots \vec{x_{N}})$ was constructed from the vectors $\vec{x_{i}}$. Then the covariance matrix $A=X\times X^{T}$ of the dimension $d\times d$ was found and its eigen values and eigen vectors were calculated. Then the set is projected on the basis of eigen vectors\cite{16}. Let us consider the phase portraits for two sounds 'a' pronounced by the same speaker in the different fragments of the speech. The sets are constructed in 3D space but for clearer comparison they are depicted in the projection on the same plane (fig. 3). 
\begin{figure}
\center{
\includegraphics[width=180pt]{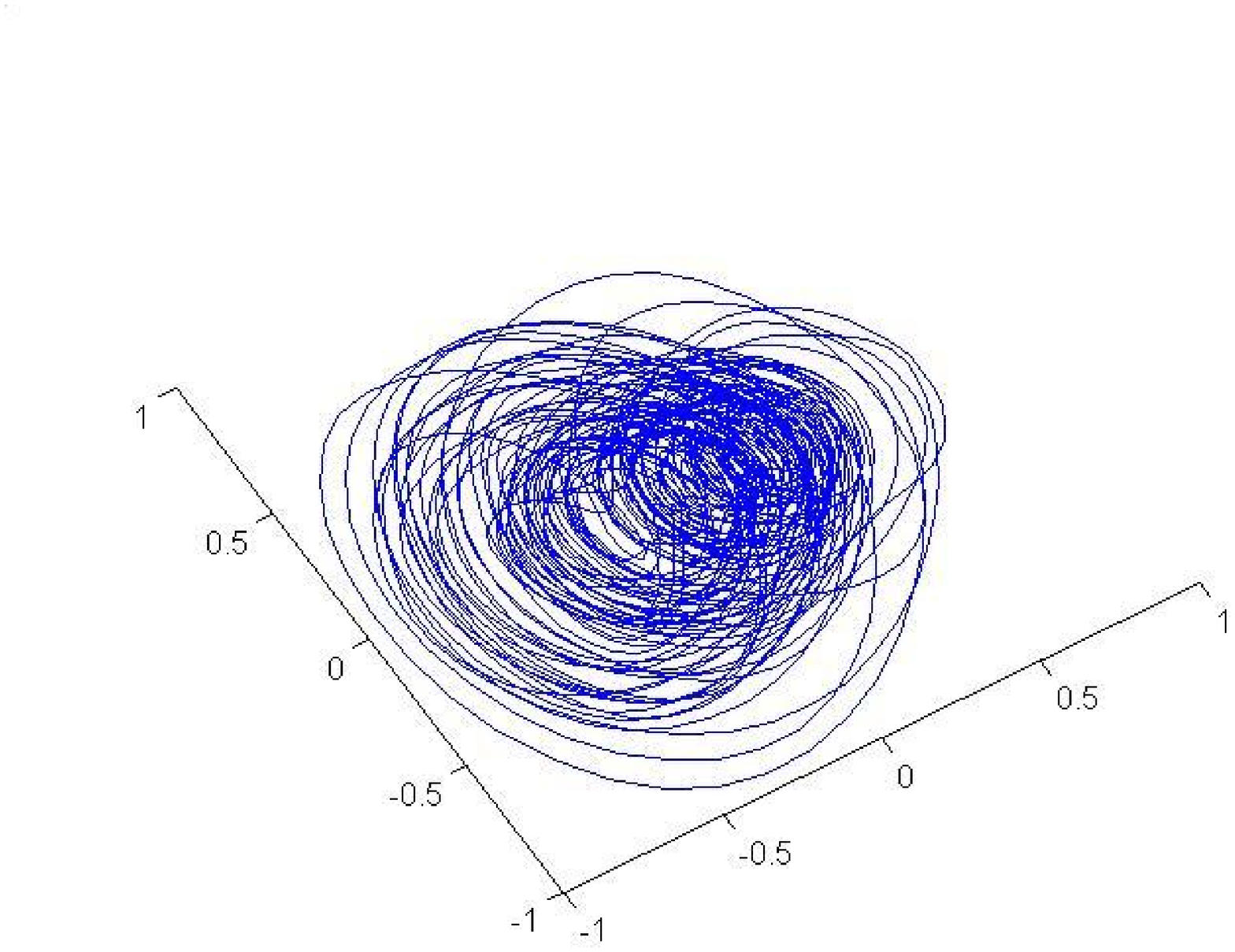}
\includegraphics[width=180pt]{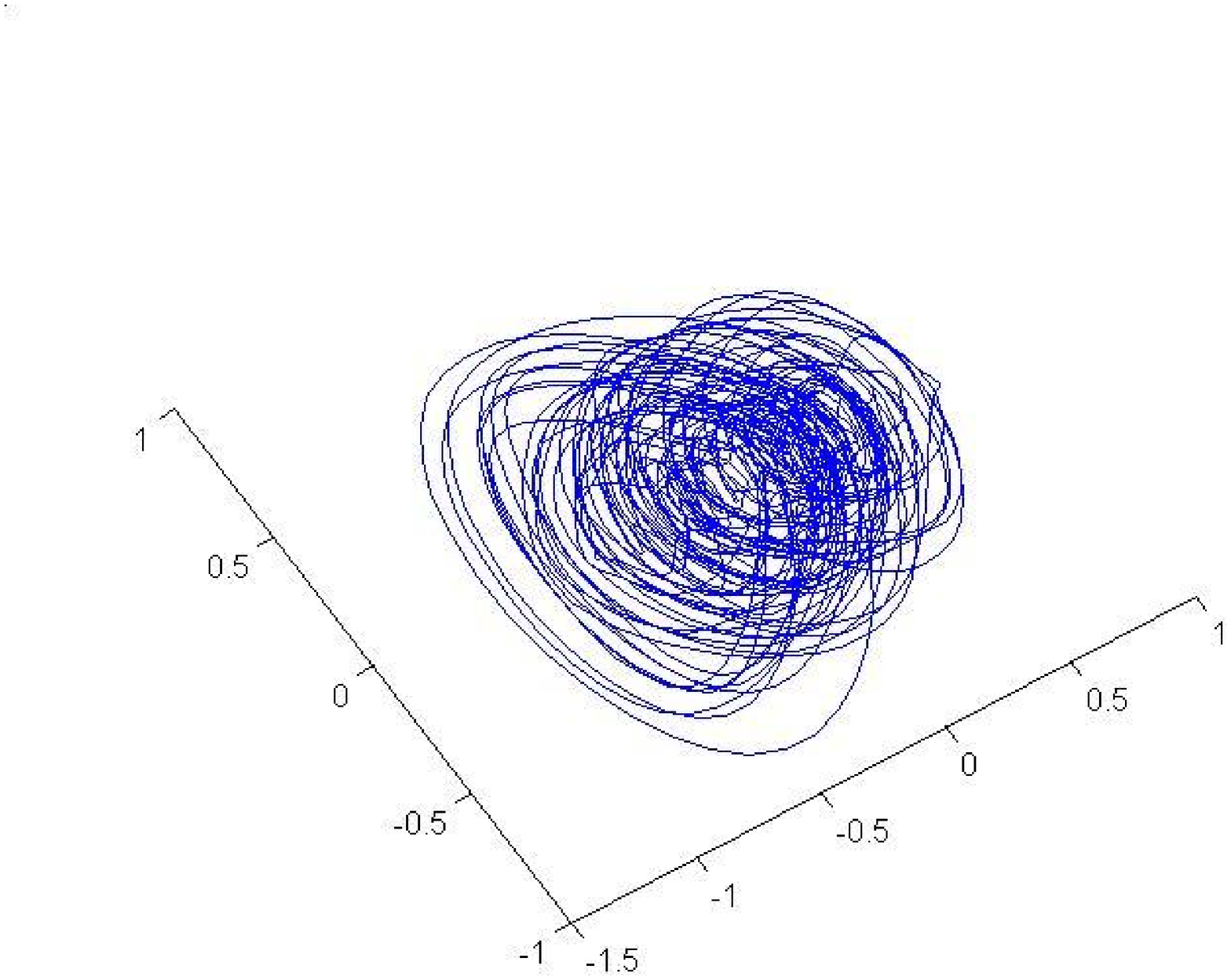}}
\caption{The phase portraits of 'a' from the real speech. The same speaker for both graphs.}
\end{figure}
In addition, the figures of the normalized eigen values which determine the dimension of the space are shown in fig. 4.
The last graph shows that the main part of energy is connected to the first and the second eigen values while the energy level corresponding to the other eigen values is significantly lower. It is seen that the fourth eigen value is of the order of $10^{-2}$. Thus, to within $10^{-2}$ the phase space can be treated as 3D one determined by the largest three eigen values.
\begin{figure}
\center{
\includegraphics[width=180pt]{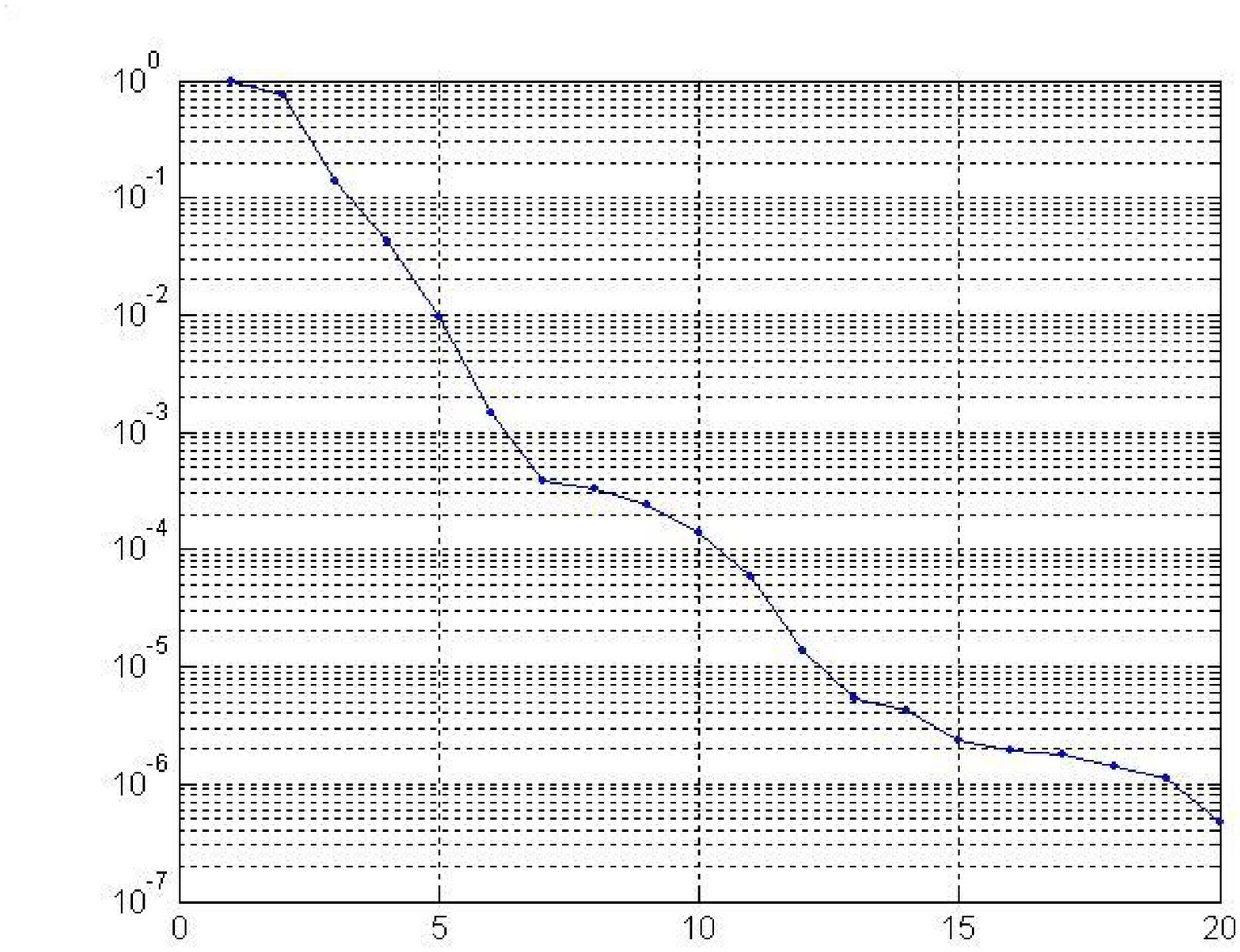}
\includegraphics[width=180pt]{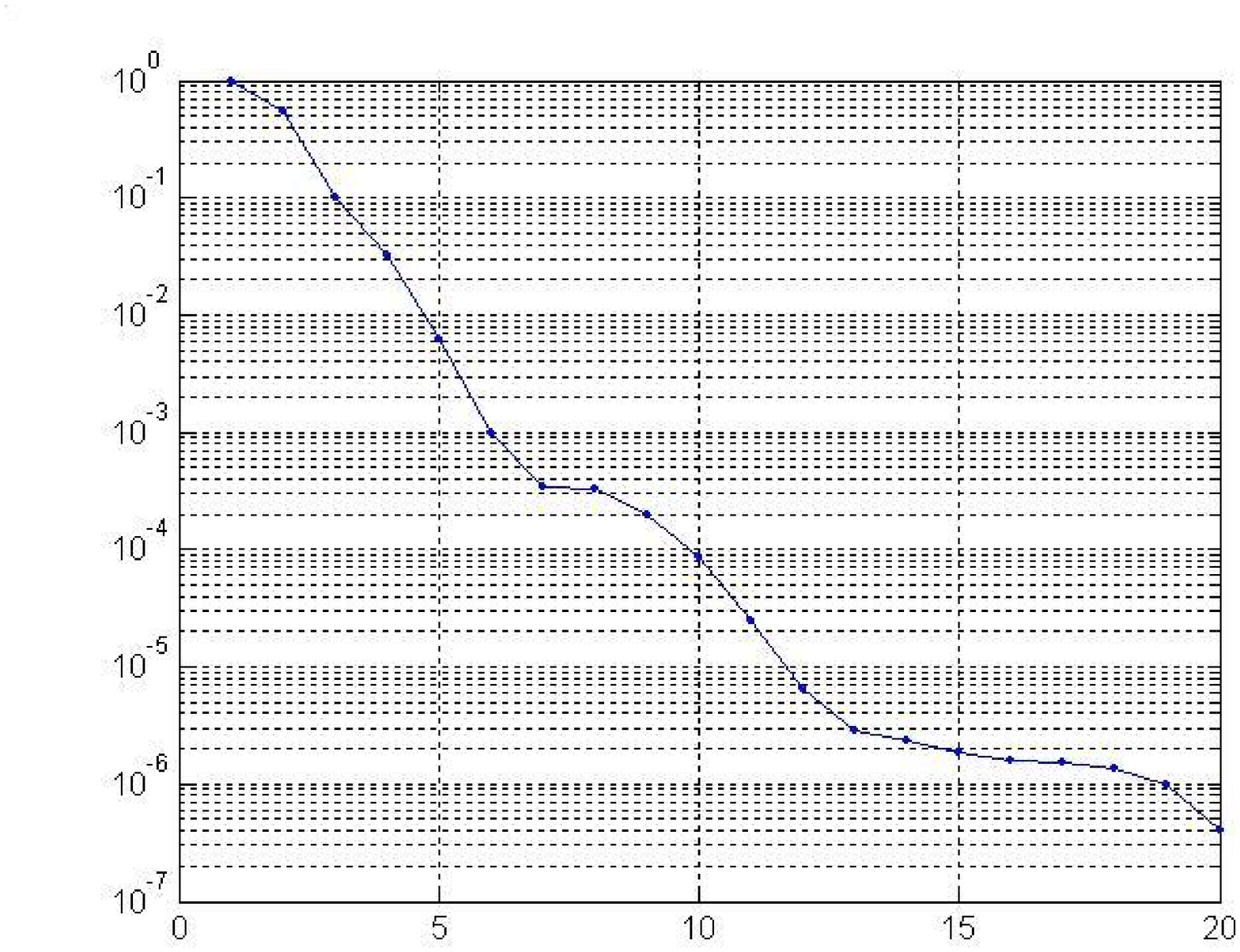}}
\caption{eigen values sorted in decreasing order for the sets in fig. 3}
\end{figure}

In general, the phase portraits of sounds pronounced by the same speaker reveal some similarity that is also valid for the other vowels. Besides, as it is seen from fig. 3, the phase portraits manifest significant smoothness implying the presence of only the low frequency components that is valid for the vowels. The slight difference in the dynamics of the same sound can be explained by the reduction of 'a' in the different fragments according to the difference of the phonemes of the sound 'a' for the speaker.

The analogous consideration was done for the other vowels. We may say that the phase portraits of vowels have the similar peculiarities for the same speaker. In addition, the structure of of the phase portraits confirms almost 'periodic' behavior of vowels, complicated by  the non-stationarity. The phase portraits of vowels display the smoothness and periodicity. The estimates of dimension for the subspace of the phase space obtained by means of the spectrum of covariance matrix show that the dimensions of the sets under consideration is not large and 
equals usually 3-4. It is noted that for all vowels there is the region of increased power in the spectrum at the level 9-13 kHz and eliminating this region slightly influence on the quality of the speech. According to the example of the vowel 'a' in the spectra of the other vowels the fundamental and formant frequencies are distinctly observed. Considering the separately pronounced vowels it was also noted that they are more stationary and have more smooth phase portraits compared to the sounds in the real speech.

\section{Dual properties of the sound 'z'}

The investigation of the voiced fricative consonant 'z' exposed some features bringing it closer both with the vowels, and with noisy hissing sounds. The phase portrait of 'z' (female voice) is shown in fig. 5a. Since this sound is not vowel it might expect to observe the significant difference of its phase portrait structure from that of vowels. However, the phase portrait has many similar features with them. In particular, the toroidal structure is quite noticeable in it. In addition, the high frequency irregularity is observed indicating the high frequency noisy component in the signal. The latter is distinctly observed in the spectrum of the signal (fig. 6). The spectral density is 30 dB lower compared to main low frequency component. However, this component of the signal seizes the band about 2000 Hz and the narrowband low frequency regular component only 10 Hz. Therefore, their integral powers differ approximately ten times. It allows to observe the noise component of the signal in the phase portraits. The structure analysis of the phase portrait and the spectrum of 'z', pronounced by female, shows that the frequency 11 kHz is apparently insufficient for its digitizing. Consequently, the male signal of 'z'  was digitized with the frequency 44,1 kHz. The typical phase portrait of this signal is depicted in fig. 5b. The sharp bends of this sound imply that even the frequency 44,1 kHz is insufficient for obtaining of the smooth data. The spectral structure of the male 'z'  differs from that for female sound (fig. 6), i.e., the spectrum contains many harmonics of the main frequency. As the spectrum of female voice the male one contains the noise component in the region from 3 to 12 kHz. Nevertheless, the line spectral components spread up to 'noise'  region of the spectrum.
\begin{figure}
\center{
\includegraphics[width=180pt]{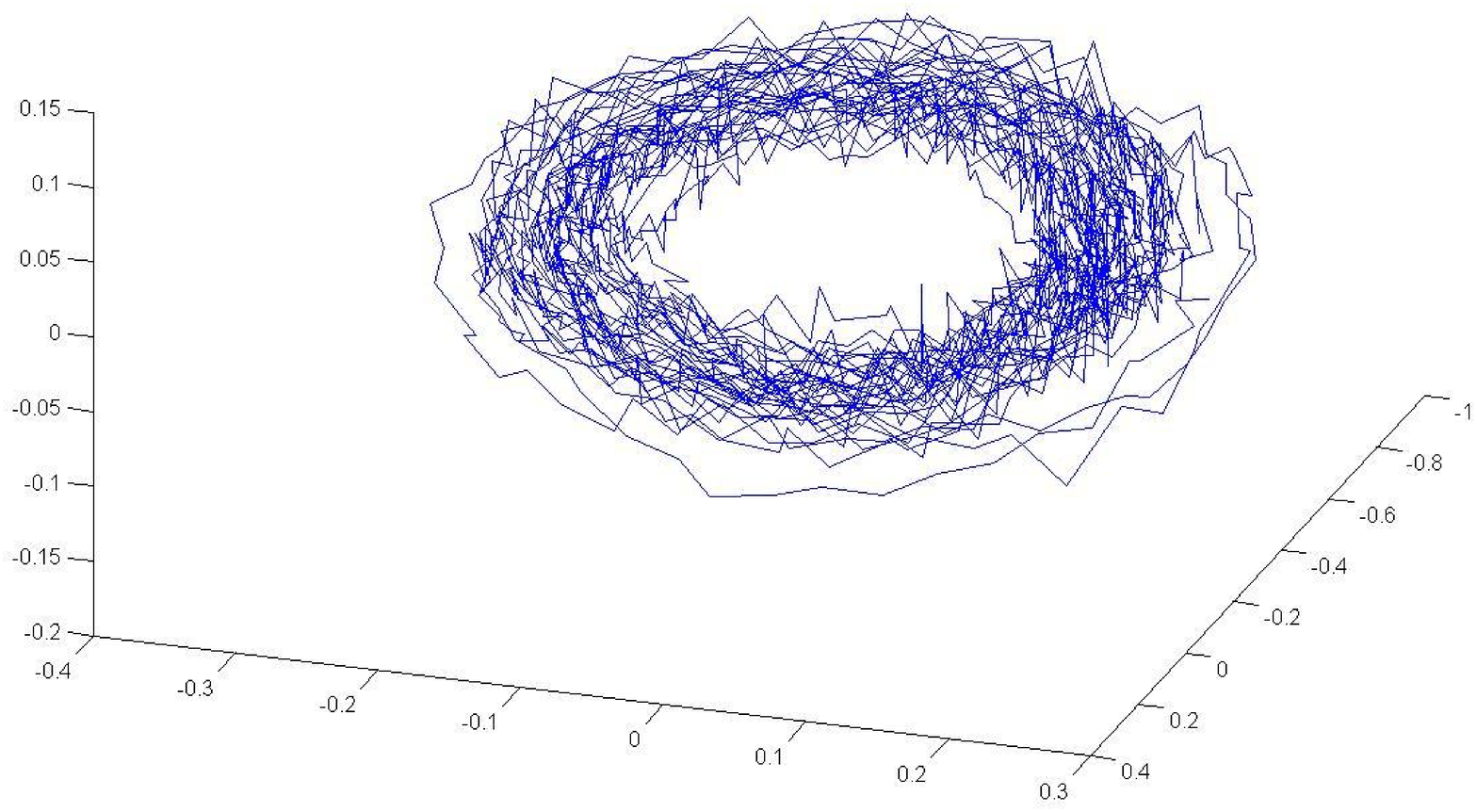}
\includegraphics[width=180pt]{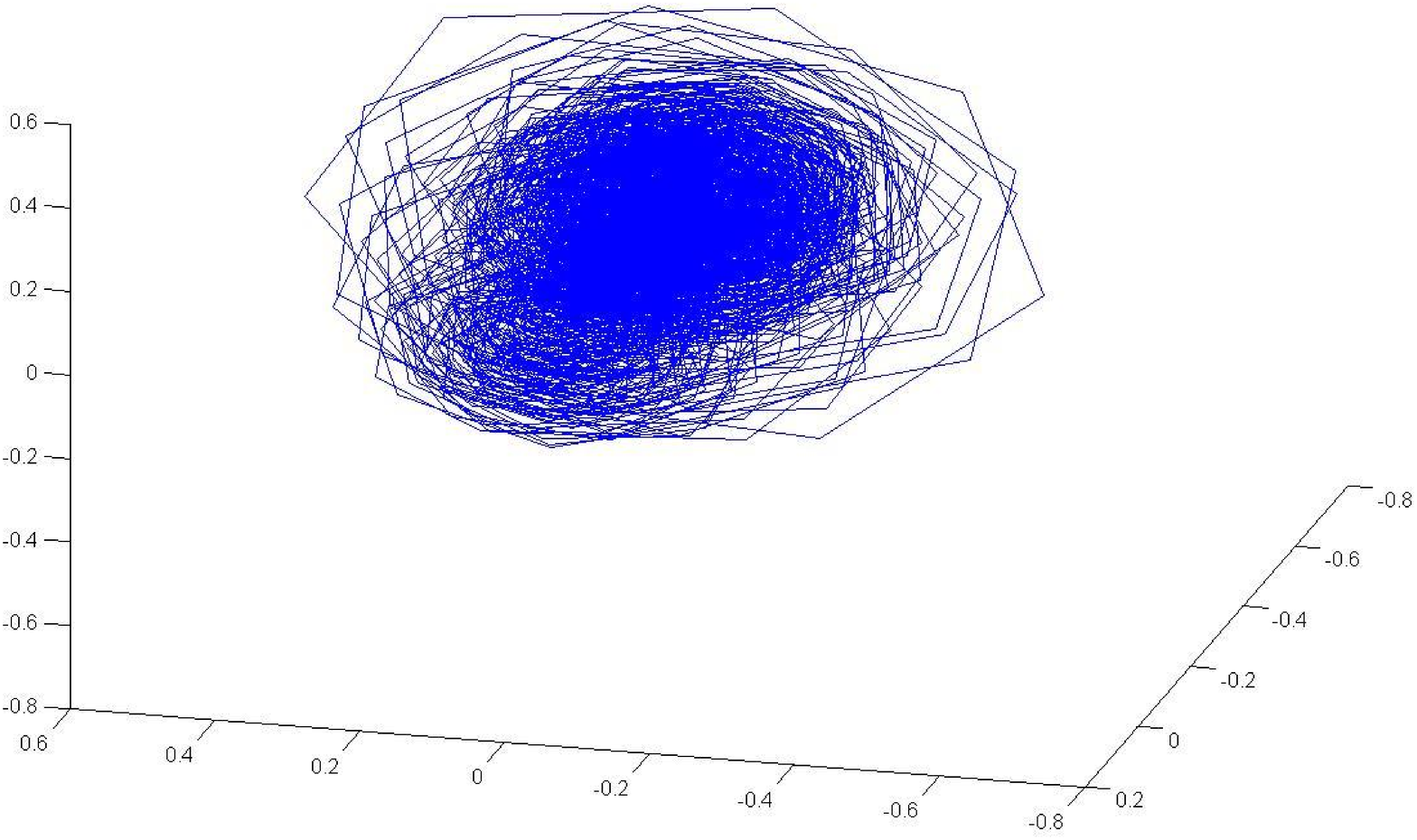}}
\caption{The phase portrait for the main components of the sound 'z'. Female and mail voices.}
\end{figure}
\begin{figure}
\center{
\includegraphics[width=180pt]{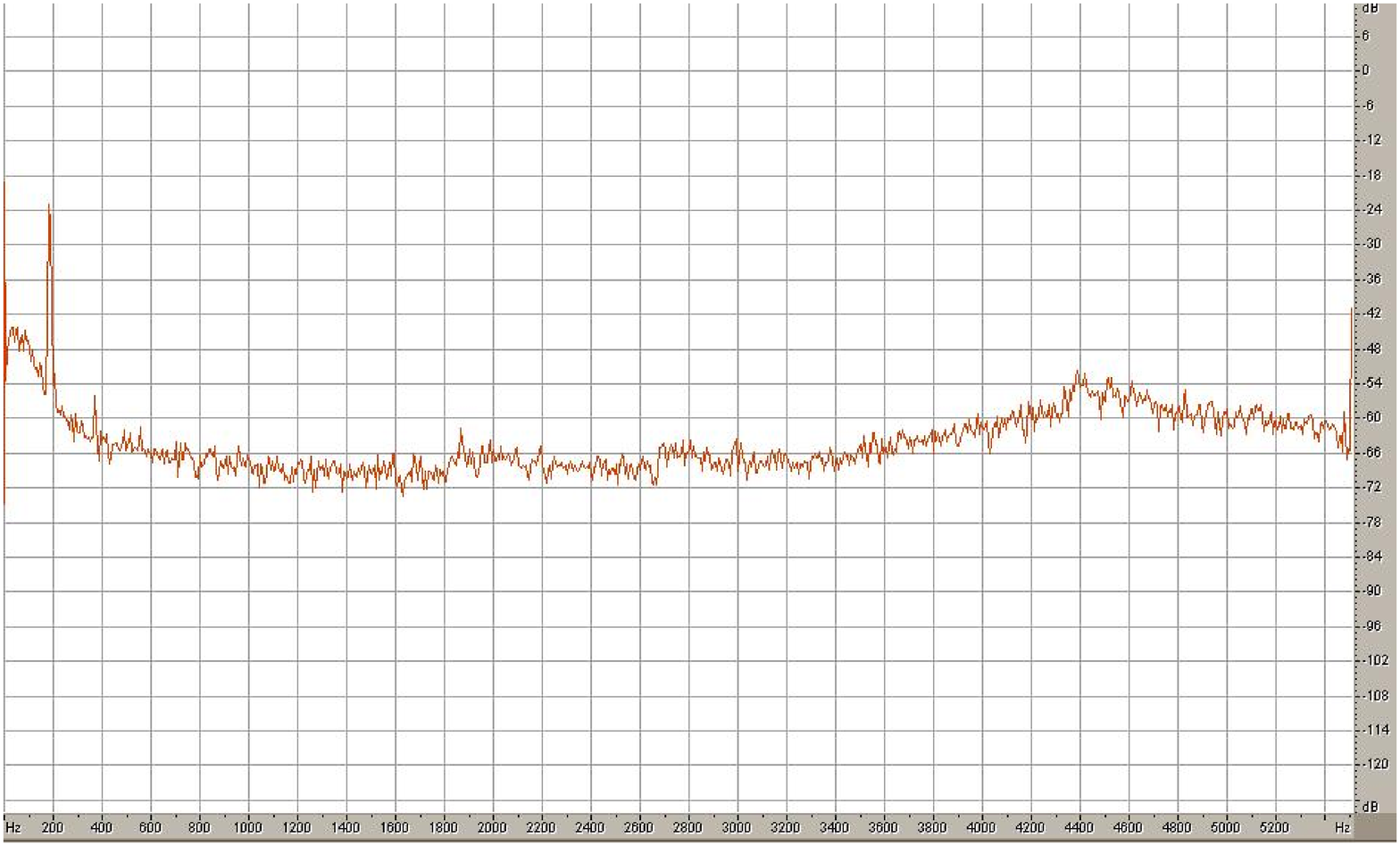}
\includegraphics[width=180pt]{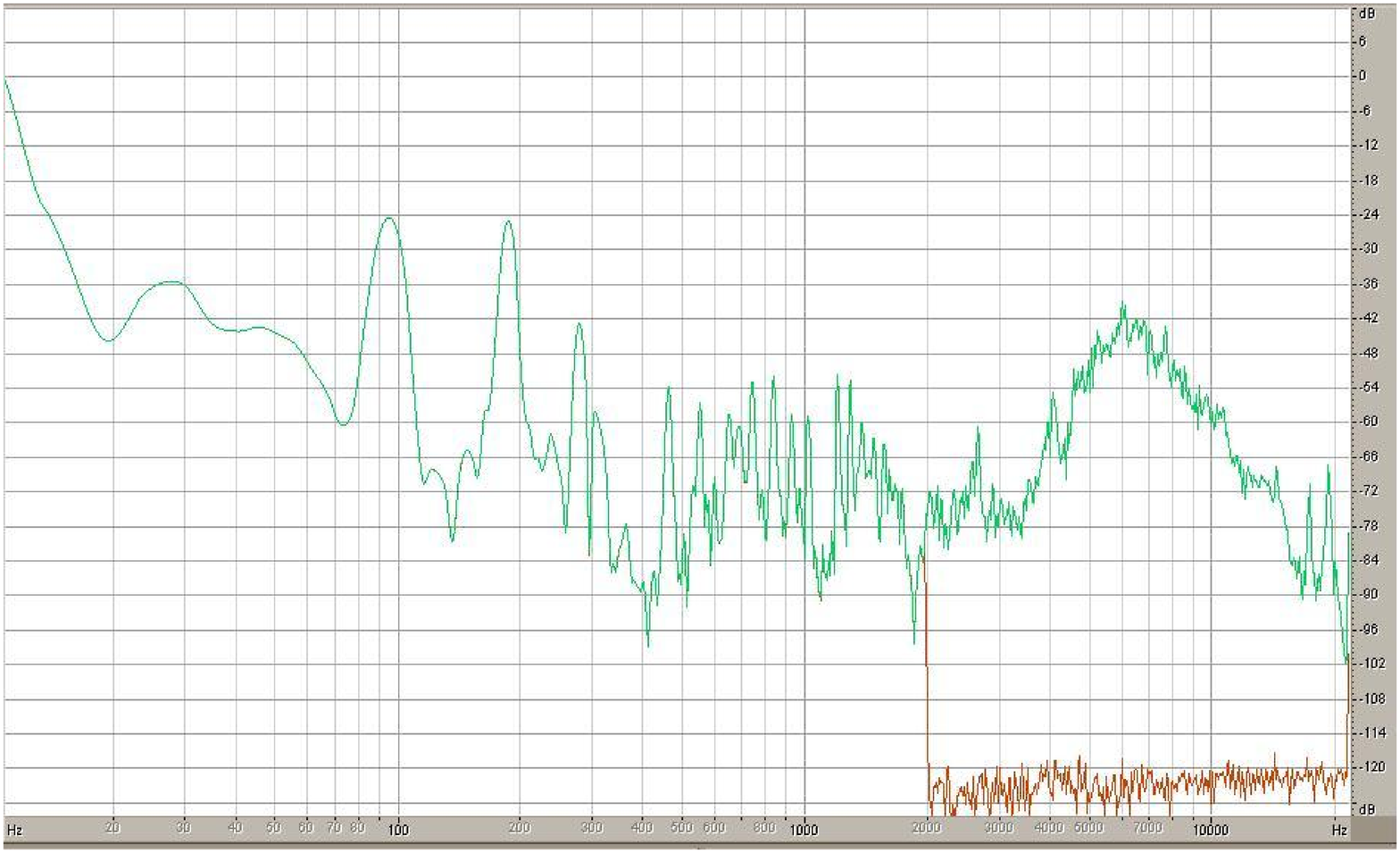}}
\caption{spectra corresponding to the phase portraits in fig. 5}
\end{figure}
Thus the sound 'z' has dual structure, it unifies from the one hand, the properties of vowels being revealed in toroidal structure of the phase portraits and in the spectrum consisting of the distinct main frequency and harmonics and from the other hand that of noiselike signals resulting in the noisy high frequency component both in the spectrum and on the phase portraits. To prove the last statement let us extract the low frequency and high frequency components from the signal. Applying the low-pass filter with the band 0-2000 Hz we obtain the signal depicted in fig. 7(above). 
\begin{figure}
\center
\includegraphics[width=200pt]{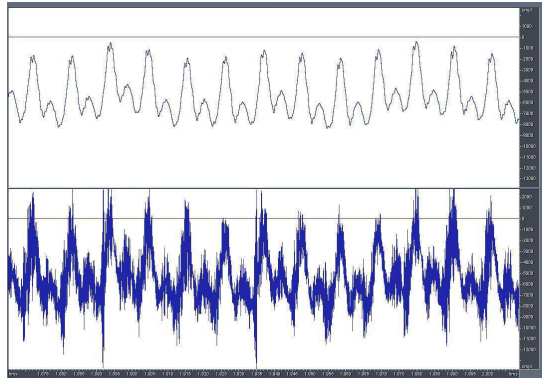}
\caption{The sound 'z'. Male voice. Above: infiltered by the low-pass filter 2 kHz, below: the original signal}
\end{figure}
The spectrum of the signal is shown in fig. 6b (the curve below). Even visually the signal is similar to that of vowels. Indeed, such a signal is heard as a voiced sound similar to 'u' or 'eu'"\footnote{This sound is absent in English}. The phase portraits of low frequency component of 'z' shows that the signal involved has the distinct toroidal structure of quite low dimension. The projection of the portrait on the coordinates 4-5-6 (fig. 8) is represented as a noisy remainder of the main and has much less amplitude. The low dimension of the obtained set is confirmed by the calculations of the eigen values spectrum of the covariance matrix resulting in the estimate $d=3$.
\begin{figure}
\center{
\includegraphics[width=180pt]{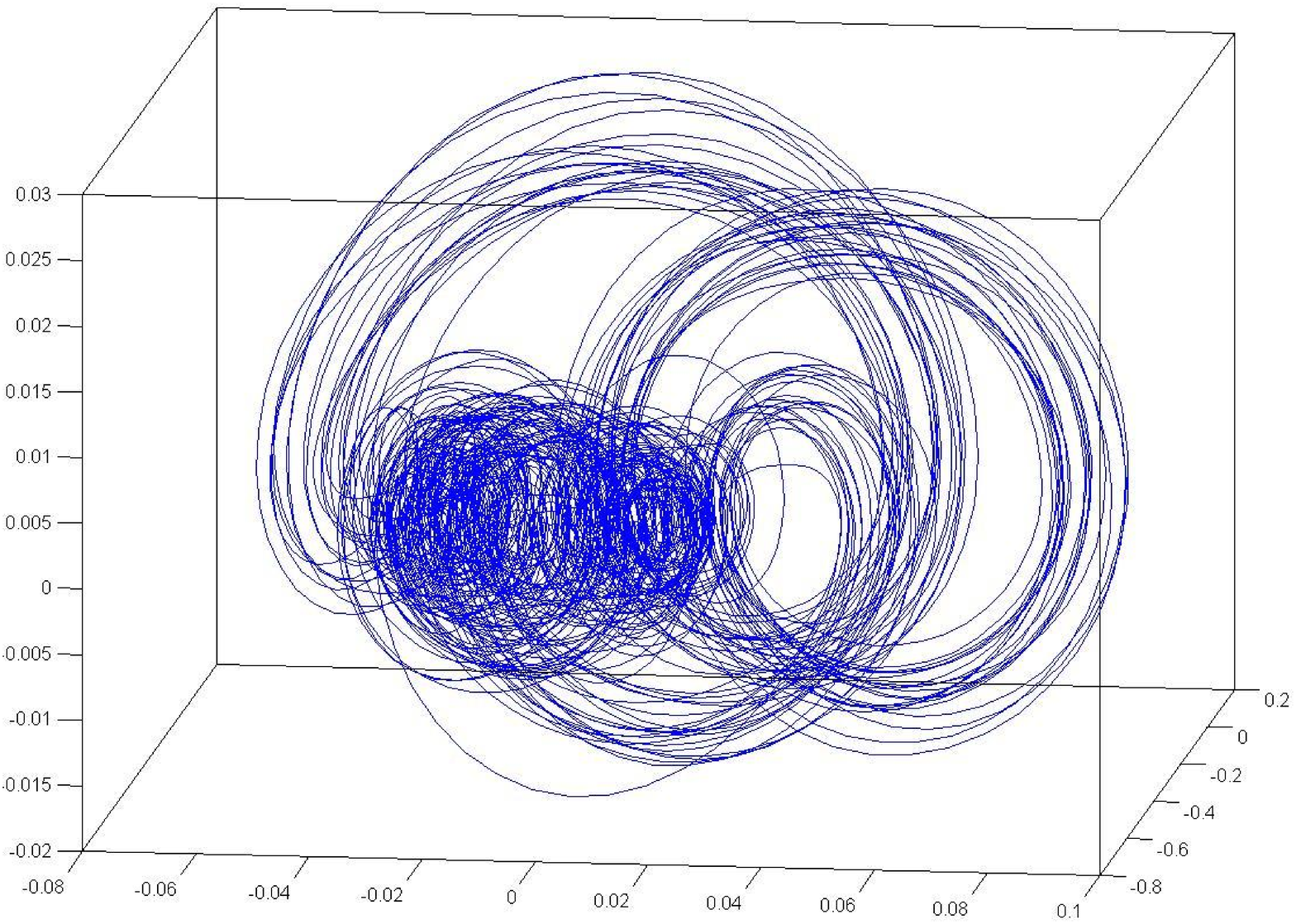}
\includegraphics[width=180pt]{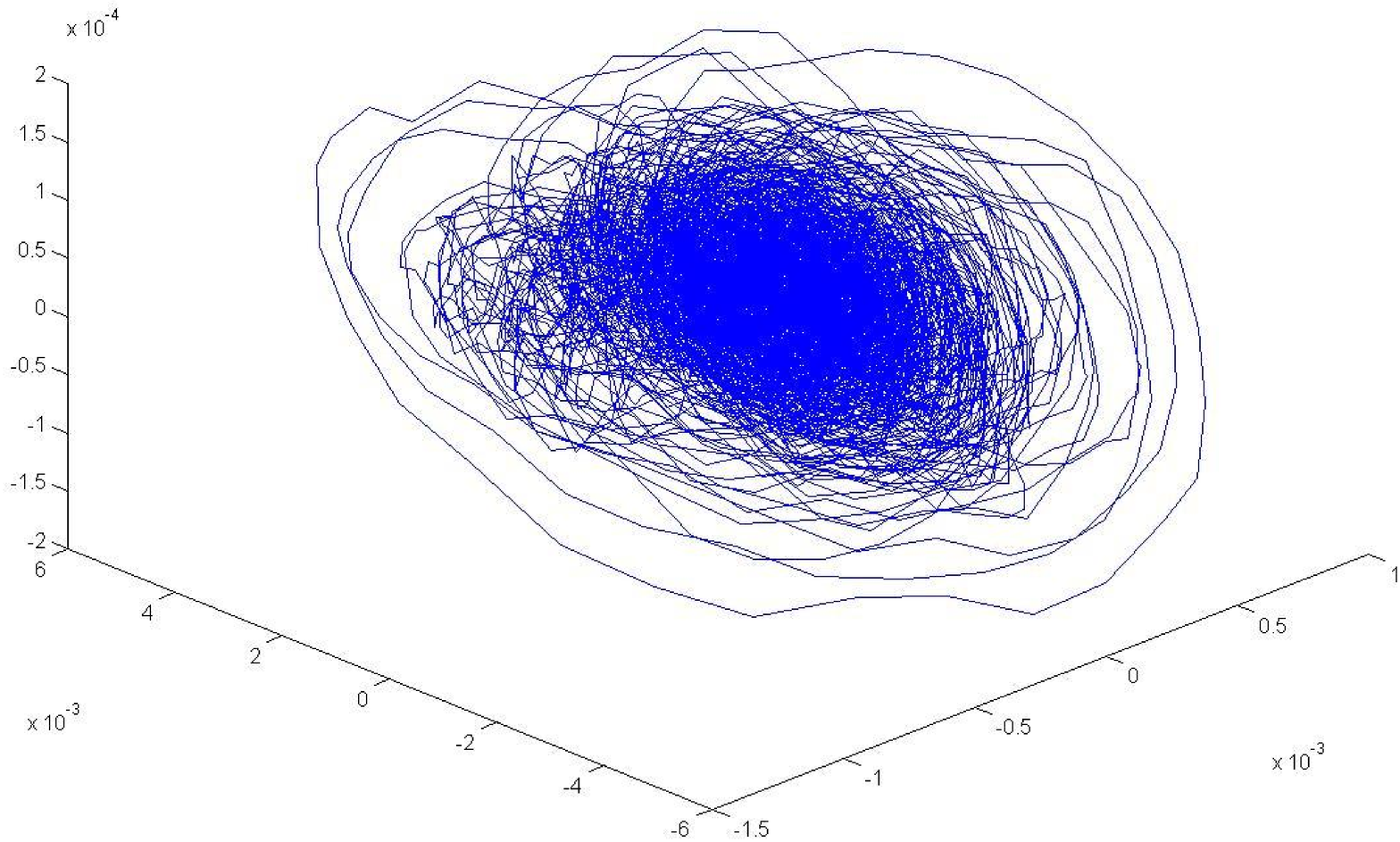}}
\caption{Projections of the phase portrait in the basis of eigen vectors of infiltered sound 'z' on the coordinates 1-2-3(left) and 4-5-6(right)}
\end{figure}
To explain the origin of the high frequency noisy region in the spectrum of the signal involved, the latter was infiltrated by means of a pass filter 3-10 kHz. The results of infiltration are shown in fig. 9.
\begin{figure}
\center{
\includegraphics[width=180pt]{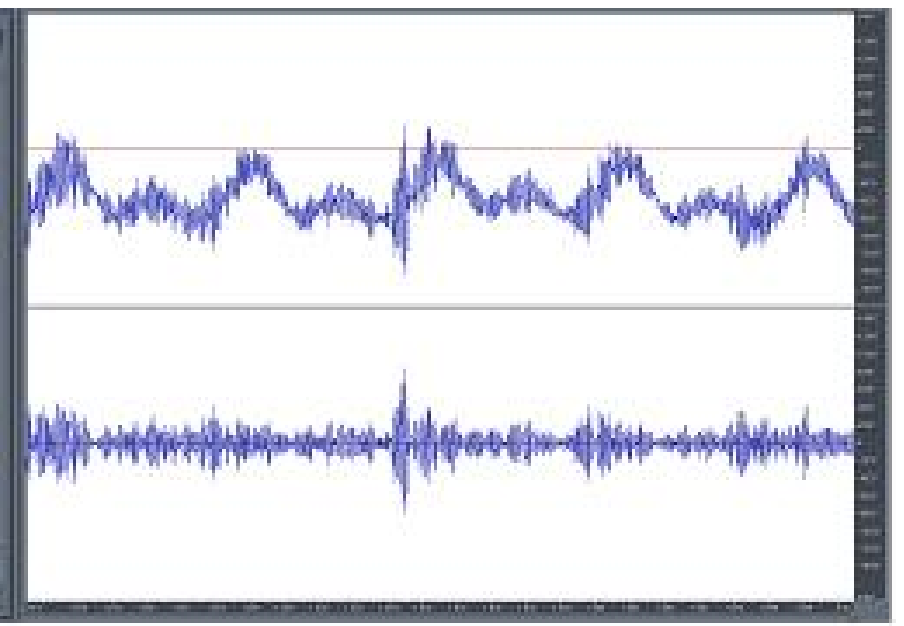}
\includegraphics[width=180pt]{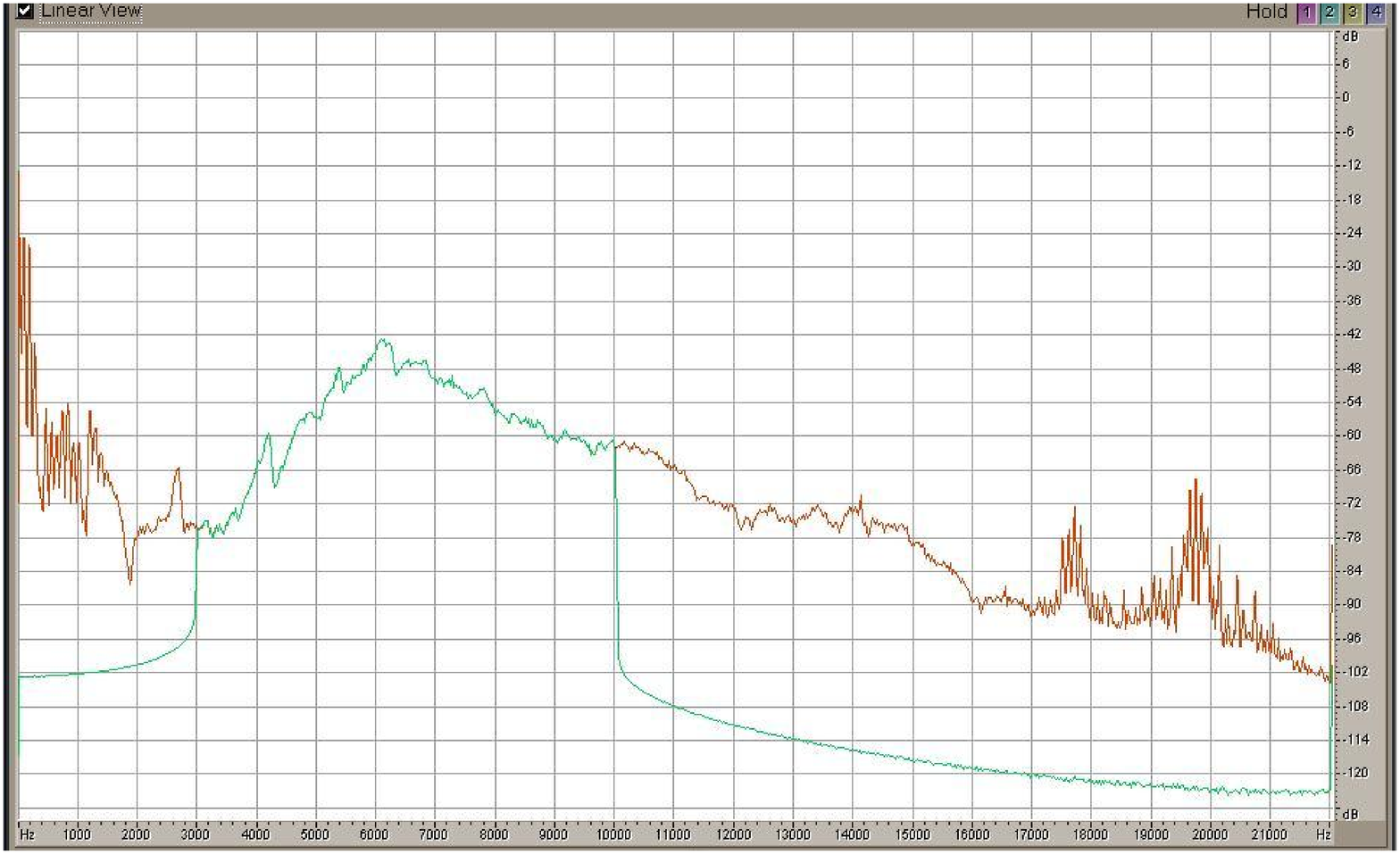}}
\caption{Characteristics of noisy part of 'z': Left graph: above - wave form of the original signal, below - infiltrated signal in the band 3-10 kHz. Right graph: the spectrum of the original (red) and infiltrated (green) signal.}
\end{figure}

By ear the infiltrated signal is perceived as hissing one, similar to 's'. According to the analysis of the eigen values spectrum the dimension of the noisy component of the signal 'z' is also restricted by the value $d=5-6$ being confirmed by the projections of the phase portrait(fig. 10). These phenomena confirm the dynamic origin of this component.
\begin{figure}
\center{
\includegraphics[width=180pt]{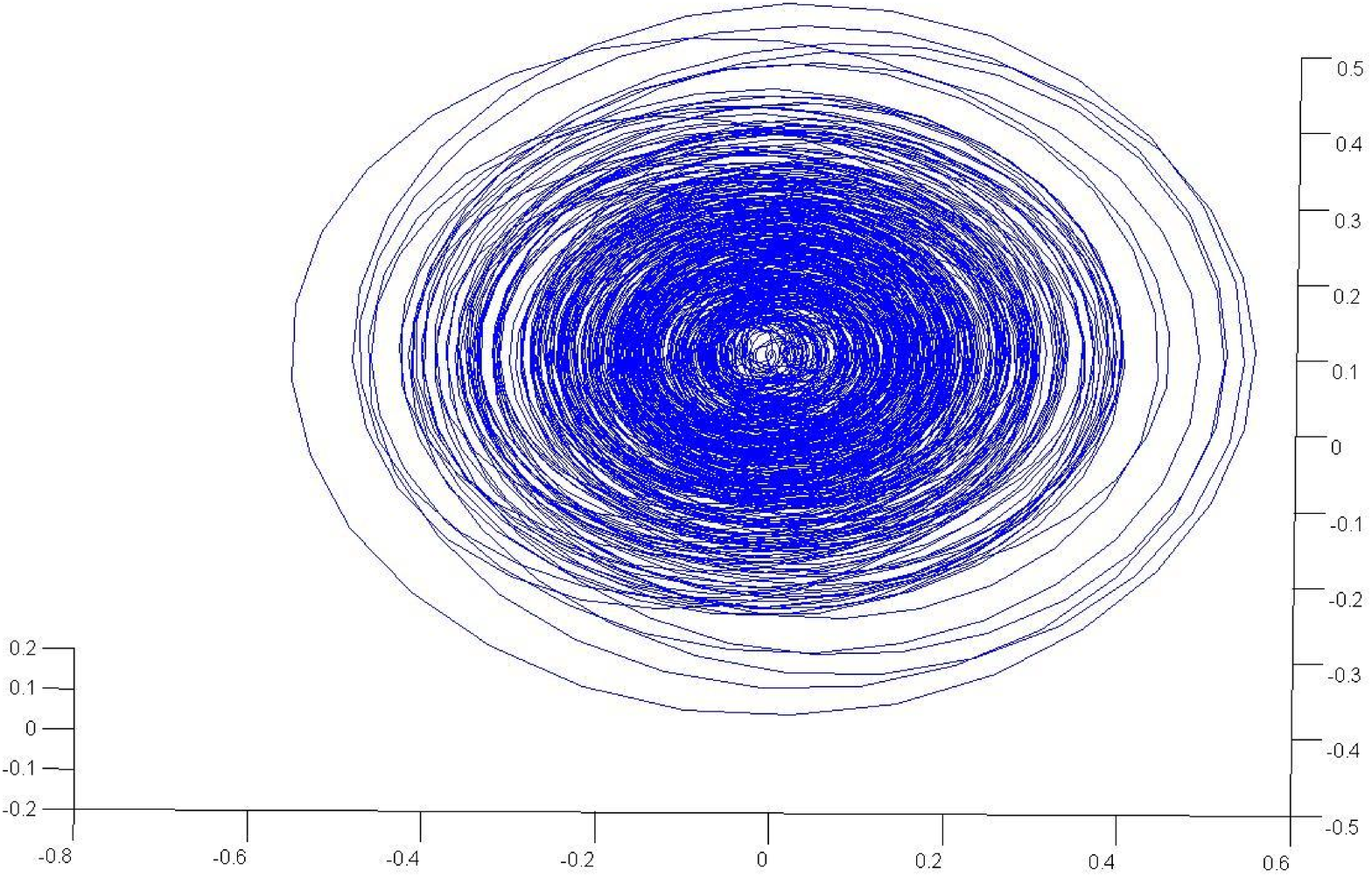}
\includegraphics[width=180pt]{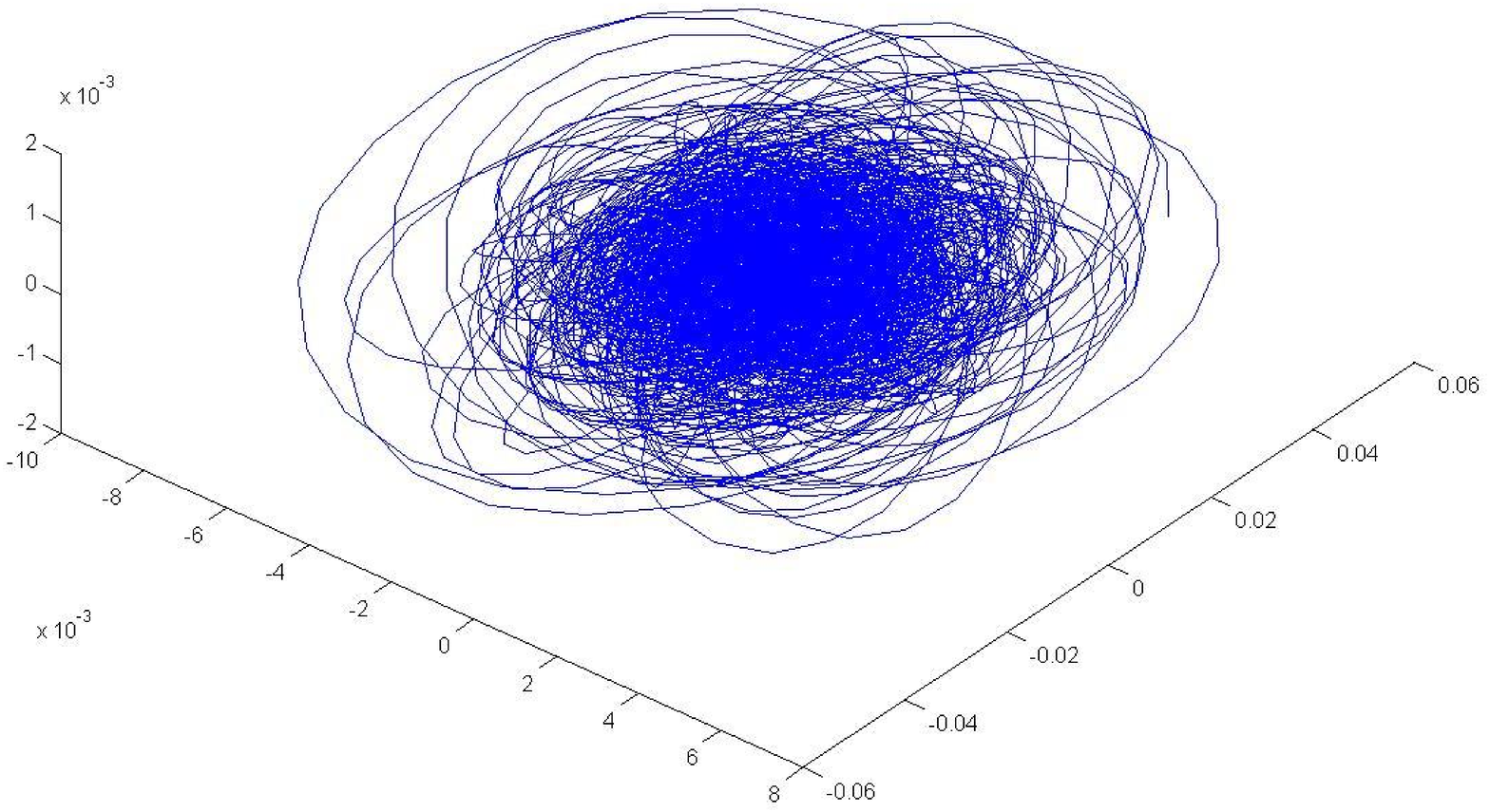}}
\caption{The phase portrait of infiltrated signal(3-10 kHz) in the basis of eigen vectors, left: coordinates 1-2-3, right: coordinates 4-5-6}
\end{figure}
Thus, the sound 'z' unifies the features of the vowel and the consonant and both components are apparently of low dimension. However, it is not clear how this sound with dual properties is formed. The authors assume that the signal in fig. 6 can be generated by means of an excitation of a nonlinear high frequency resonator by a comparatively low frequency harmonic signal but the hypothesis requires to be investigated.  

\section{Estimation of the largest Lyapunov exponent}

It is well known that $D$-dimensional nonlinear system is characterized by the set of $D$ Lyapunov exponents. If the system is in chaotic mode then at least the largest Lyapunov exponent is positive. It follows that two trajectories leaving a small neighborhood diverge exponentially. In the simple systems typically the largest Lyapunov exponent is positive, the others are negative. In our experiments we estimated the largest Lyapunov exponent for various sounds.

This estimate is not complicated if the equations of the system are given. In this case for the points on the trajectory of the system the local coefficients of stretching are estimated which then average resulting in the estimate of the largest Lyapunov exponent.
To obtain the local coefficient of stretching in the time moment $t_{k}$ in the neighborhood of the current point $x_{k}$, a close point $x^{\prime}_{k}$ is taken and the trajectory on the interval $\Delta t$ going from $x_{k}$ is calculated. The local Lyapunov coefficient is estimated as follows 
$$
\lambda_{k} = \frac{1}{\Delta t}\log\left(\frac{\norm{x_{k+1} - x^{\prime}_{k+1}}}{\norm{x_{k} - x^{\prime}_{k}}}\right),
$$
where $x^{\prime}_{k+1}$ and $x_{k+1}$ - points of appropriate trajectories in the time $t_{k} + \Delta t$. Then, the vector $x^{\prime}_{k+1} - x_{k+1}$ is normalized the length to be small and the procedure is repeated. 

If the governing equations are unknown but only the observable trajectory of the system is given, then the estimate of the trajectory close to the observable is impossible. In this case the local linear predictors are applied for estimating the trajectories, adjacent to the observable\cite{17}. If the latter is sufficiently long then it supposed to run many times in the same region of the phase space and in the neighborhood of any point there would be many other points. First, the vectors in the neighborhood of the current point have to be found. Then the linear transform is constructed mapping the neighborhood of the current point to that of the consequent point on the trajectory. Using the matrix of the obtained linear transform, the stretching coefficients of deviation vectors on the time interval of transition from the current point to the next one are evaluated. The logarithm of this coefficient represents the local Lyapunov exponent in the current point. In the neighborhood of the next point the procedure is repeated. The Lyapunov exponent of the dynamic system is then evaluated by means of averaging the local exponents on a long interval of the trajectory\cite{18,19}. It is evident that the precision of the estimate depends on linearity of the transform of the neighborhood to the following one, i.e. is determined by the sufficiently large number of the points in the neighborhood of every point on the trajectory, which is, in turn, depends on the length of the trajectory.

Let us give the estimates of the largest Lyapunov exponent for a vowel. Below we represent the results for the other sounds as well but a more detailed description we give for the vowel 'a'. In fig. 11 the fragment and the reconstructed phase portrait for the sound 'a' are shown. In the figure of the signal two time scales are easily viewed, 'large period' $T_{1}$ 50 samples, corresponding to the fundamental frequency 200 Hz (discretizing frequency 11025 Hz) and the 'period' of small oscillations $T_{2}=13$ (850 Hz). The phase portrait of the signal in the embedding space is shown in the right figure. The portrait is depicted by two originally very close trajectories (the initial deviation 0,024). Having diverged, these two trajectories represent a complex toroidal portrait of the signal.
\begin{figure}
\center{
\includegraphics[width=180pt]{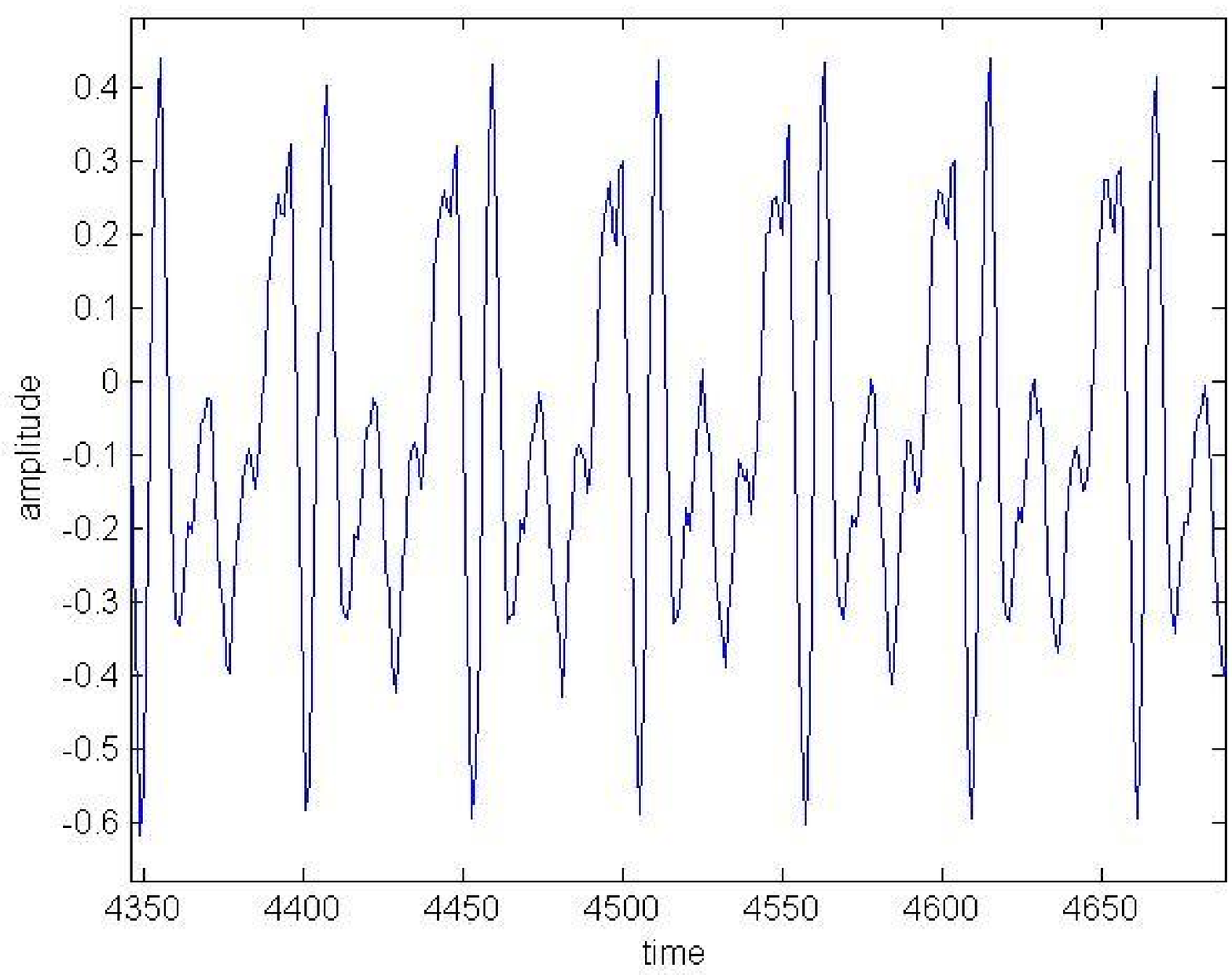}
\includegraphics[width=180pt]{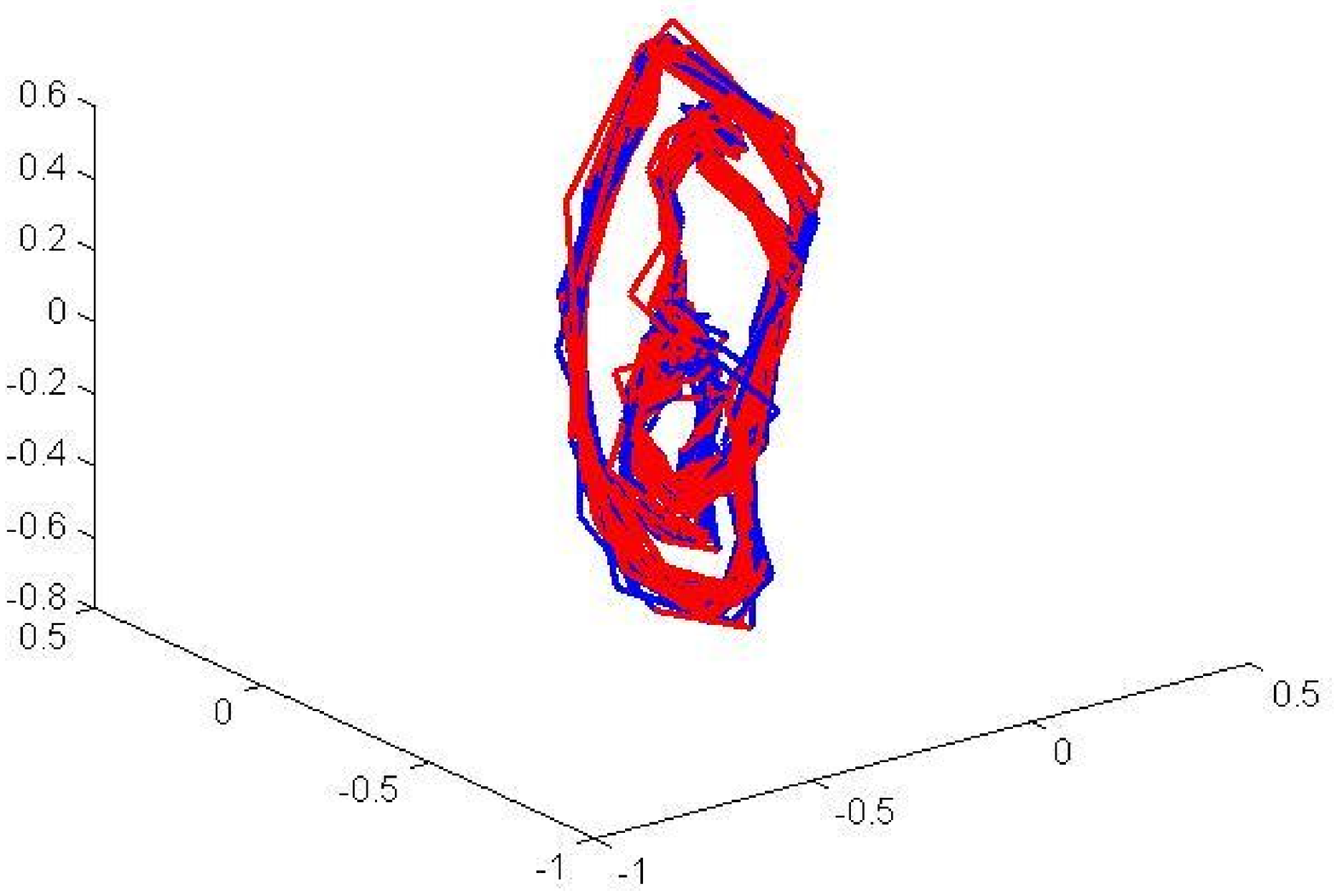}}
\caption{The wave form of 'a' sound and its phase portrait. Blue line - the original trajectory, the red line - locally linear predictor}
\end{figure}
The largest Lyapunov exponent estimating by means of the above method gives $\lambda\sim 600^{-1}$ seconds. 
\begin{center}
\begin{tabular}{|l|r|c|}
\hline
\multicolumn{3}{|c|}{The largest Lyapunov exponent for the sounds}\\
\hline
sound & $\lambda$ & F=11 kHz, N of samples \\
\hline\hline
e(f) & 5050(8,4) &  10000 \\
\hline
a(m) & 2830(4,7) & -\\
\hline
u(f) & 4750(7,9) & 20000 \\
\hline
u(m) & 732(1,2) & -\\
\hline
ae(f) & 3210(5,3) & 10000\\
\hline
ae(m) & 3800(6,3) & 10000\\
\hline
o(f) & 2750(4,5) & 10000\\
\hline
o(m) & 2160(3,6) & 10000\\
\hline
s(f) & 6360(10,6) & 10000\\
\hline
sh(f) & 6140(10,2) & 20000\\
\hline
sh(m) & 2830(4,7) & 20000\\
\hline
\end{tabular}
\end{center}
To compare Lyapunov exponents of the real signals to model systems such as, e.g., Lorenz system, they have to be transformed to the same time scale. In the Lorenz system the period of small oscillations equals $\sim 0.7$ but in our case $\sim$ 13/11026 i.e., the scales differ approximately 600 times. On the scale of Lorenz system for the sound 'a' we obtain $\lambda\sim 1$, i.e., the Lyapunov exponents for both systems are comparable. For the vowels 'u' and 'e' the similar values were found on the scale of Lorenz system. However, it is essential to notice the following fact. The positive value of the largest Lyapunov exponent implies the chaotic behavior inherent in the speech dynamics. In the table the results of evaluations for the largest Lyapunov exponent are represented for various vowels and consonants. It should be noted that the signals are different from the stationarity viewpoint. We give the absolute values of the exponent and for comparison give the values reestimated according to the time scale of Lorenz system. It is noted the significant increase of the Laypunov exponent for the consonants. In addition, as it was mentioned above, the values of the exponent depend on the length of the trajectory fragment because for the longer intervals the non-stationarity effects turn out to be crucial.

\section{Estimation of correlation dimension \\of speech signals}

One of the main method for estimating the dimension of the set obtained in the embedding space for the speech signal is that of correlation dimension estimate. The problem is quite similar to that when the dimension of the chaotic attractor is evaluated. To estimate the correlation dimension, first, the correlation integral is calculated\cite{6,14}
$$
C(\epsilon, N) = \frac{1}{N^{2}}\sum_{i\ne j}\theta(\epsilon - \abs{\vec{x}_{i}-\vec{x}_{j}}),
$$
where $\vec{x}_{i},\vec{x}_{j}$ are the points of $d$ dimensional embedded space and $\theta$ - Heaviside function. Then the correlation dimension is determined from
$$
\nu = \lim_{\epsilon\to 0}\lim_{N\to\infty}\frac{\log C(\epsilon,N)}{\log\epsilon},
$$
where $N$ the number of the points in the embedded space\cite{20}. The correlation dimension also gives the estimate of the set for the speech signals in the phase space\cite{13,20}. We show the results of the correlation dimension estimations for various vowels and consonants pronounced by various speakers and for small fragments of the speech. In the fig. 12 and 13 the wave forms and correlation dimension graphs are displayed for the various dimensions of the embedded space. 
\begin{figure}
\center{
\includegraphics[width=180pt]{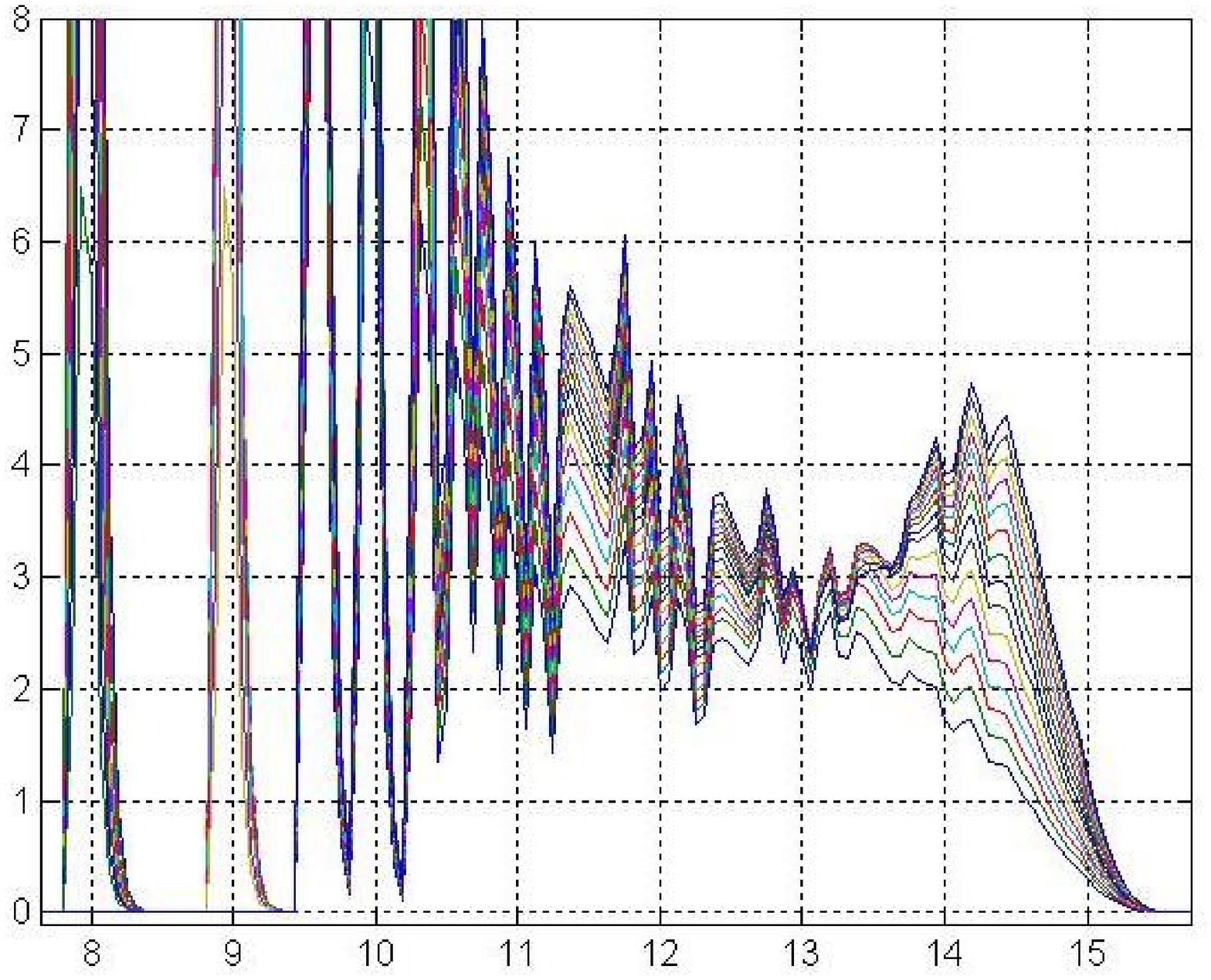}
\includegraphics[width=180pt]{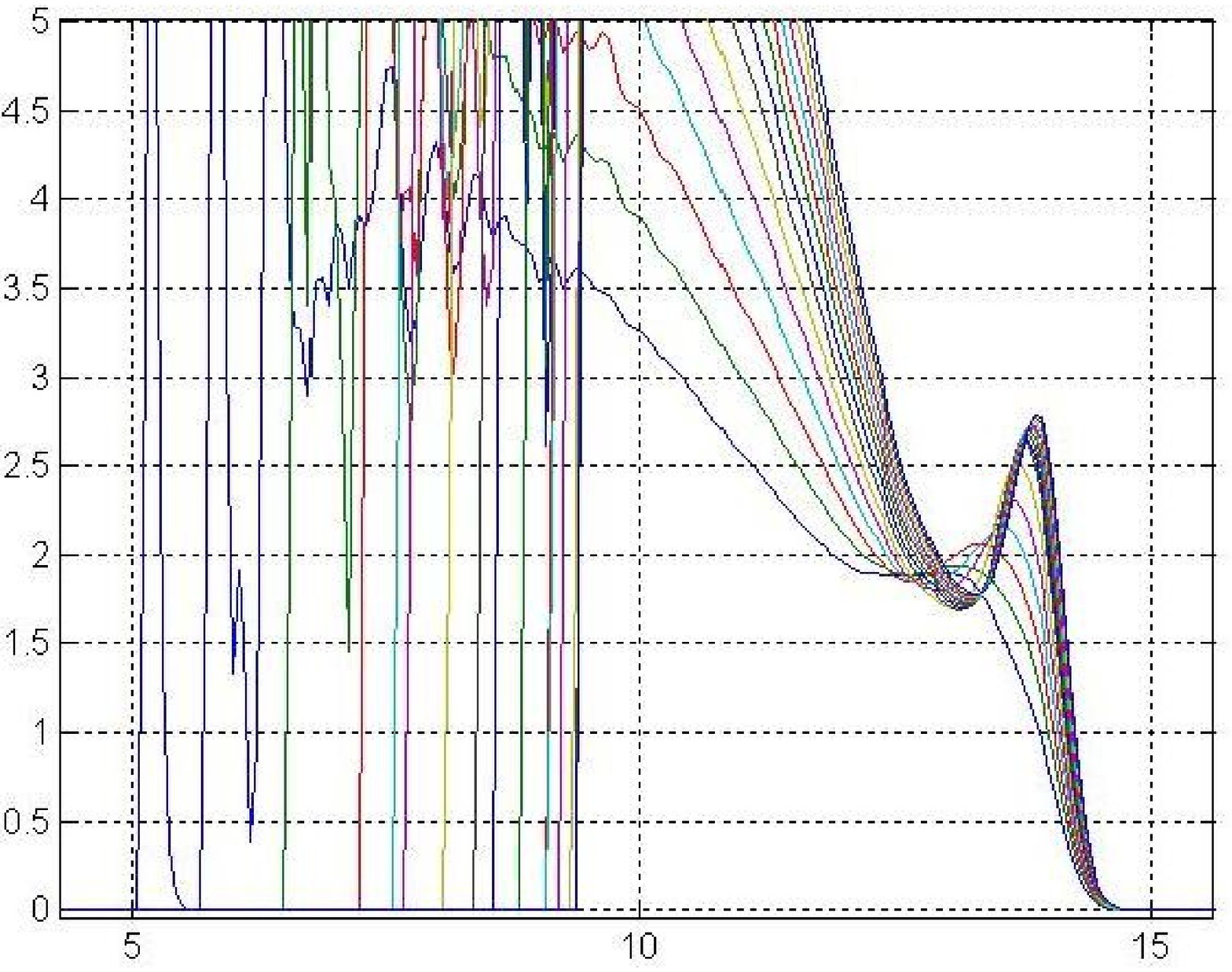}}
\caption{Correlation dimensions of 'a' and 'u' vowels. Male voice. Embedding dimensions=4-18.}
\end{figure}
\begin{figure}
\center{
\includegraphics[width=180pt]{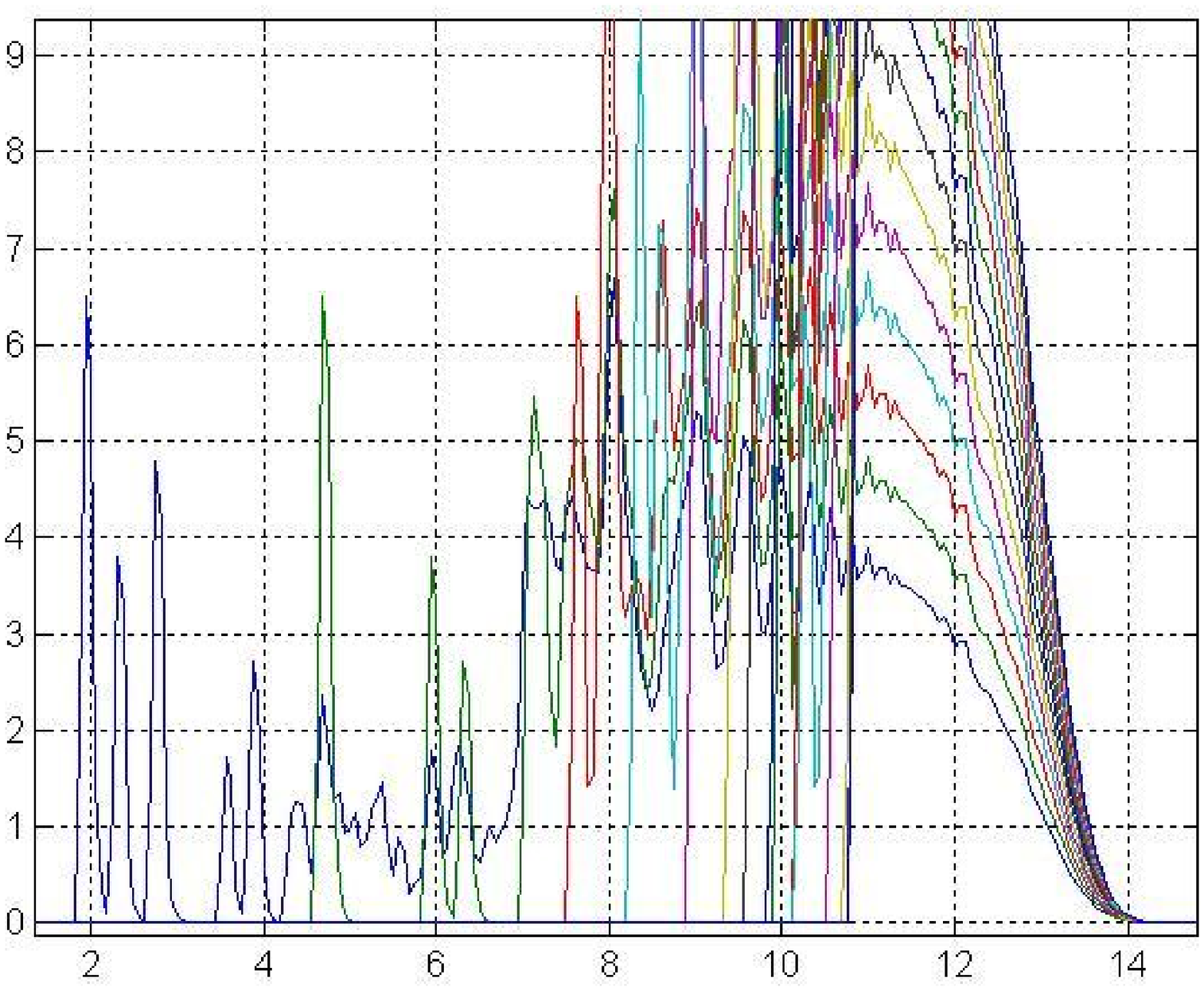}
\includegraphics[width=180pt]{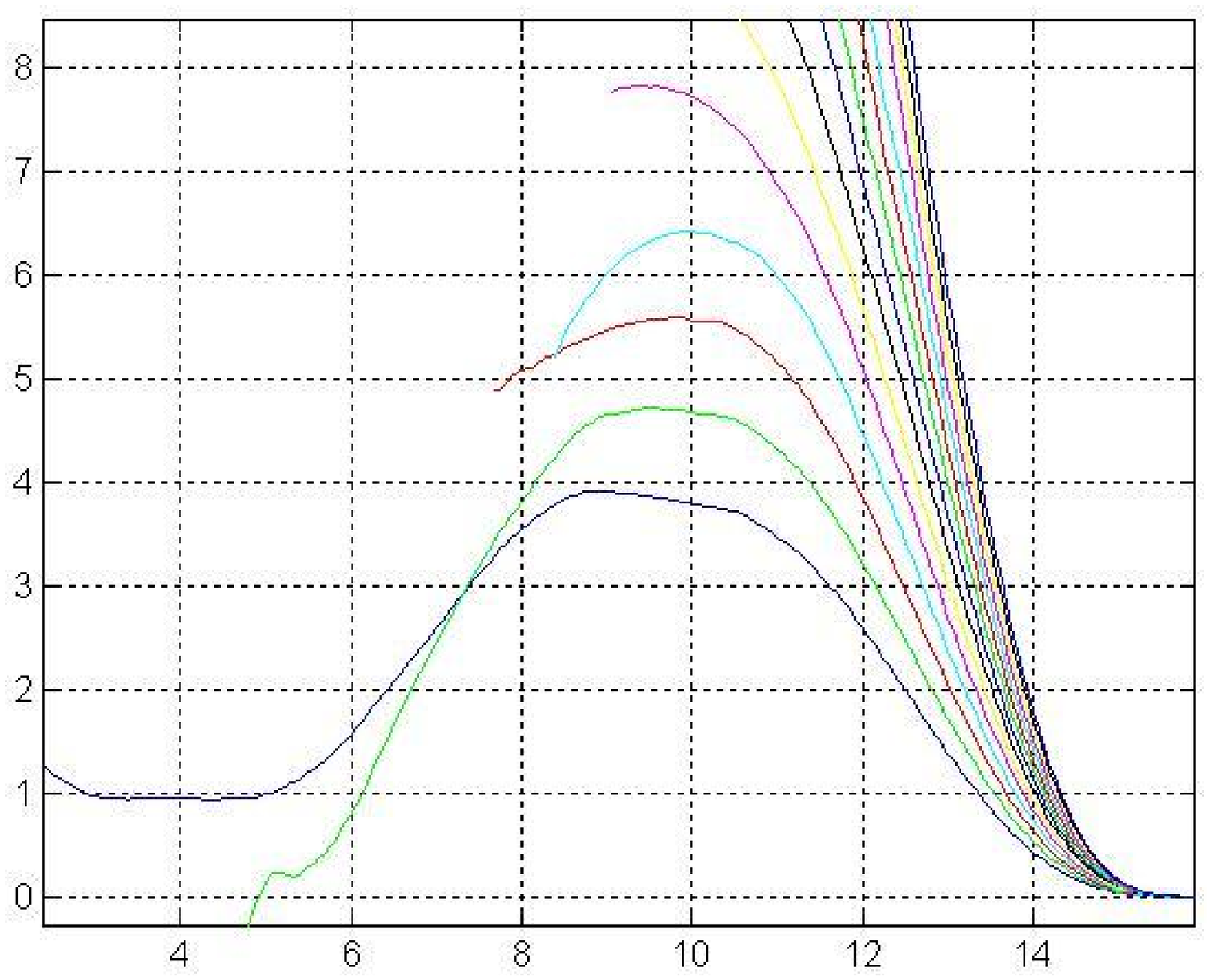}}
\caption{Correlation dimensions of 'sh'. Female voice. Embedding dimensions=4-18. Right graph shows the smoothed curves.}
\end{figure}
The estimate of correlation dimension gives a significant result only if the growth of the embedded space dimension results in some saturation, i.e., when the curves corresponding to the different dimensions of the space begin to approach each other. This effect is seen in the fig. 12 and the saturation region gives the estimate of the dimension 2,5-3 in both cases. The noisylike sound 'sh' results in another effect. For it the significant interval of saturation was not obtained implying the increase of the dimension of the set in this case. Nevertheless, it should be noticed that this fact conforms with the phase portraits for that sound which revealed that the energy is distributed on the eigen values more uniformly than in the case of vowels.

\section{Conclusion}

In this work the results of applications of nonlinear dynamics methods for the speech analysis have been represented. We show the data for various vowels and consonants both recorded for this research and taken from the real speech. Because of the huge amount of data and results we illustrate them mainly with the data obtained for the vowel 'a' and for some consonants. In the similar way we analyzed the vowels 'e', 'ae', 'o'. We deliberately postponed the analysis of diphthongs because of their more complicated structure. All vowels taken from the real speech have quasi-periodic structure in contrast to the consonants which are in part of noisy properties and in part have to be considered as complicated transition processes. In general, we would like to notice the following. While studying the speech signals we deal with a complex non-stationary system generating the speech signal and the chaotic behavior in the system. The main purpose of this work consisted in detection of nonlinear dynamic nature of the system of vocal tract. This problem, in our opinion, can not be solved in one publication being extremely vast one. We restricted ourselves only with analysis of some sounds, yet from this sounds one should transmit to their combinations and then, to the real speech. It was shown that when embedding according to the method by Takens in a space of larger dimension, the vowels can be defined in it as sets located in a subspace of smaller dimension. For all the sounds the dimension of the space equals about 3-4. The represented results of the sound analysis confirm the dominance of nonlinear effects in speech signal forming.

The peculiarities of sounds pronouncing intrinsic to the speaker can be detected, considering the reconstructed sets in the phase space of the dynamic system, as a specific structure inherent in the sets. Thus, it was noted that the peculiarities of the speech producing for a person can be treated geometrically. This fact might open the way  for collecting the banks of speech signals being treated from that viewpoint and can be applied in various technologies. The calculation of the largest Lyapunov exponent, accomplished  
for all the sounds involved and its positive value implies the chaotic behavior in the system of vocal tract being also the crucial argument for the chaotic nature of the speech.

The calculations of the correlation dimension for vowels give the estimate of the embedded space of various sounds. For all vowels this estimate is about 3-4 while for the consonants it occurred to be difficult to obtain the distinct value of this characteristic.

\end{document}